\documentclass[runningheads]{llncs}
\usepackage{graphicx}
\usepackage{comment}
\usepackage{amsmath,amssymb} 
\usepackage{xcolor}
\usepackage{booktabs}
\usepackage{multirow}
\usepackage{subcaption}
\usepackage{wrapfig}
\usepackage{hyperref}
\graphicspath{{figures/}}
\usepackage[width=102mm,left=22mm,paperwidth=146mm,height=173mm,top=22mm,paperheight=217mm]{geometry}

\definecolor{umich}{HTML}{00274c}
\definecolor{tblue}{HTML}{174992}
\definecolor{Gray}{gray}{0.93}
\newcommand{\RN}[1]{%
	\textup{\lowercase\expandafter{\it \romannumeral#1}}%
}
\begin{document}
\pagestyle{headings}
\mainmatter
\def\ECCVSubNumber{7020}  

\title{Structure-Aware Human-Action Generation} 

\titlerunning{Structure-Aware Human-Action Generation}
%
\author{Ping Yu\inst{1}, Yang Zhao\inst{1}, Chunyuan Li\inst{2}, Junsong Yuan\inst{1}  \and Changyou Chen\inst{1} }
\authorrunning{P. Yu et al.}
%
\institute{The State University of New York at Buffalo \and
Microsoft Research, Redmond \\
\email{\{pingyu, yzhao63, jsyuan, changyou\}@buffalo.edu
}~
\email{chunyl@microsoft.com
}
}
\maketitle

\begin{abstract}
Generating long-range skeleton-based human actions has been a challenging problem since small deviations of one frame can cause a malformed action sequence. Most existing methods borrow ideas from video generation, which naively treat skeleton nodes/joints as pixels of images without considering the rich inter-frame and intra-frame structure information, leading to potential distorted actions. Graph convolutional networks (GCNs) is a promising way to leverage structure information to learn structure representations. However, directly adopting GCNs to tackle such continuous action sequences both in spatial and temporal spaces is challenging as the action graph could be huge. To overcome this issue, we propose a variant of GCNs (\textbf{SA-GCNs}) to leverage the powerful self-attention mechanism to adaptively sparsify a complete action graph in the temporal space. Our method could dynamically attend to important past frames and construct a sparse graph to apply in the GCN framework, well-capturing the structure information in action sequences. 
Extensive experimental results demonstrate the superiority of our method on two standard human action datasets compared with existing methods. The code to reproduce our analysis is available at
\textcolor{tblue}{{\bf \href{https://github.com/PingYu-iris/SA-GCN}{https://github.com/PingYu-iris/SA-GCN}}}.

\end{abstract}

\section{Introduction}

Recent years have witnessed the development of skeleton-based action generation, which has been applied in a variety of applications, such as action classification \cite{hrnn2015,repactrec2017,interpre2017,shahroudy2016ntu,yan2018spatial}, action prediction \cite{hpgan2017,julieta2017motion,pred_eccv2018} and human-centric video generation \cite{wang2018vid2vid,yan2017skeleton}. Action generation is still a challenging problem since small deviations in one frame can cause confusion in the entire sequence.

One of the most successful methods for skeleton-based action generation considers skeleton-based action generation as a standard video generation problem
\cite{cai2018deep,habibie2017recurrent,wang2019learning}. Specifically, the method naively treats skeleton joints as image pixels and sequential actions as videos, without considering the rich structure information among both joints and action frames. 
The video-generation based methods may produce distorted actions when applied to skeleton generation, if prior structure knowledge is not well leveraged. A first step to consider structure information into action generation is to represent a skeleton as a graph structure to characterize the spatial relations between joints in each frame based on graph convolution networks (GCN) \cite{zhou2018graph,kipf2016semi,bruna2013spectral}. However, most existing GCN methods do not have the flexibility to process continuous sequential graphs data. 
This poses a new challenge: {\em how to construct a representation to effectively incorporate both temporal and spatial structures into action generation?}


Generally speaking, there are two classes of methods with GCN to model action structure information: 
$(\RN{1})$ \textit{Full connection}: an entire action sequence is7890-=-098765 considered as a graph. Each node of the current frame is connected with the corresponding nodes in all the past frames. This construction, however, is computationally very inefficient (if ever possible at all). Moreover, the model could be highly redundant since many frames are similar to each other. 
$(\RN{2})$ \textit{Spatial-temporal graph convolutional networks} \cite{yan2018spatial}:  a graph convolution is first applied to intra-frame skeletons, whose extracted features are then applied with a 1D convolution layer to capture temporal information. This method typically requires weight sharing among all nodes, and the ability to model temporal information is somewhat weak. 

\begin{figure}[!t]
    \centering
    \includegraphics[width=0.85\linewidth]{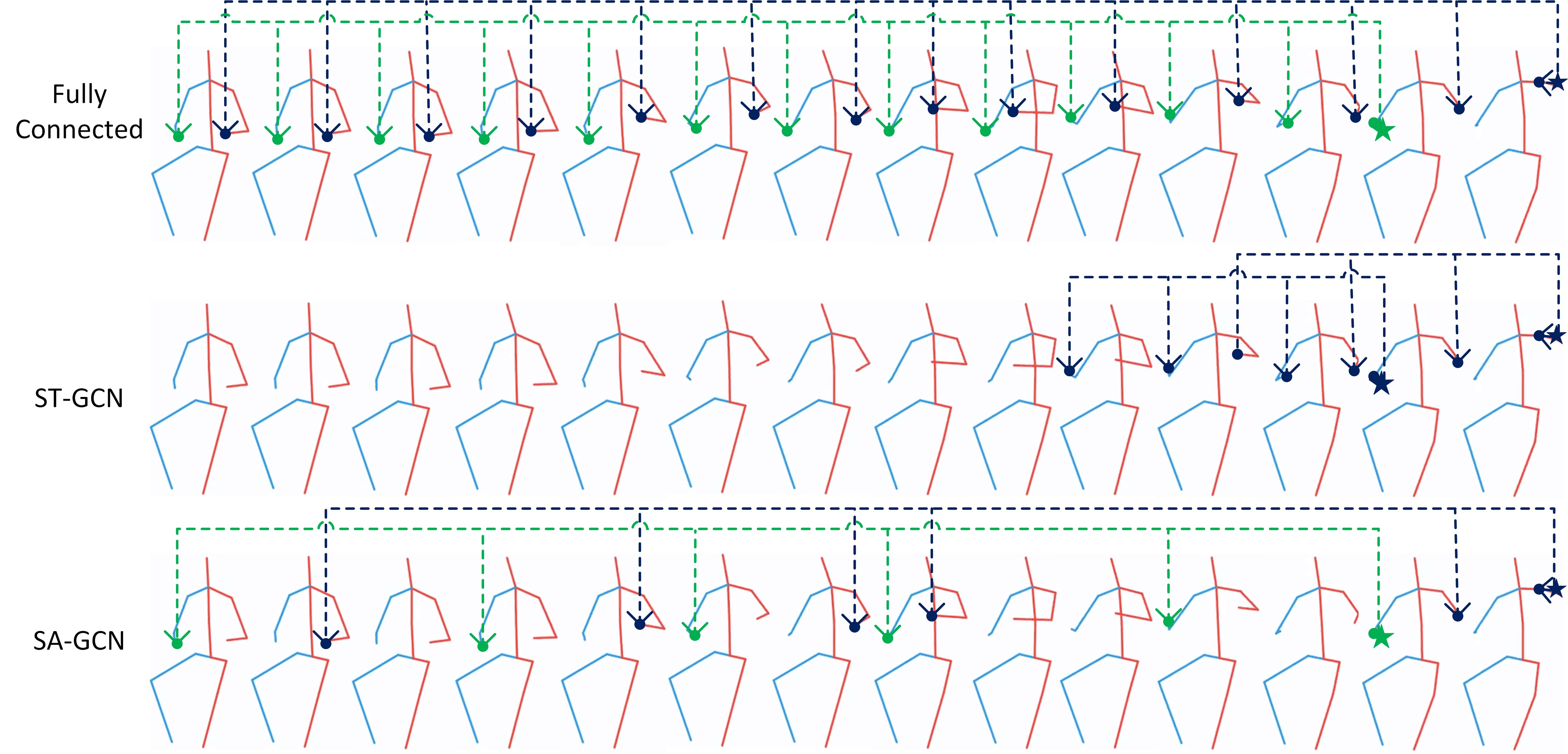}
    \caption{
    \small{Comparisons of the construction of action graphs with our proposed method (3rd tow) and two standard methods (1st and 2nd rows) to encode temporal information. First row (\textit{full connection}): the left-hand joint gather information from all left-hand joint of past frames; similar to the right-hand joint. Second row (\textit{ST-GCN}): a 1D convolution of kernel size $k$ is used to encode temporal information. Both the left and right hands could encode information from past $k$ frames with share weights. Third row (\textit{SA-GCN}): both the left- and right-hand joints learn to encode information from a selected left-hand joints based on the attention scores.} }
    \label{fig:introduction}
\end{figure}

We advocate that a better solution should be proposed to leverage skeleton structures and gather information from action sequences more efficiently. In this paper, we propose {\it Self-Attention based Graph Convolutional Networks (SA-GCN)} to build generic representations for skeleton sequences.
Our \textit{SA-GCN} aims at building a sparse global graph for each action sequence to achieve both computational efficiency and modelling efficacy. Specifically, for a given frame, the proposed {\it SA-GCN} first calculates self-attention scores for other frames. Based on the attention scores, top $k$ past frames with the most significant scores are selected to be connected to the current frame to construct inter-frame connections. Within each frame, the joints are connected as the original skeleton representation. To demonstrate the differences between our construction and the aforementioned two constructions, Fig.~\ref{fig:introduction} illustrates a sequence of samples in terms of every three consecutive frames on the Human 3.6m dataset \textit{SittingDown} sequence. 
As illustrated in the figure, our method can be considered as an adaptive scheme to construct an action graph, with each node assigning a trainable weight instead of a shared weight as in other methods. 

The major contributions of this work are summarized in three aspects:
\begin{itemize}
    \item We propose {\it SA-GC} layer, a generic graph-based formulation to encode structure information into action modelling efficiently. Our method is the first sparse and adaptive scheme to encode past frame information for action generation.
    \item By efficiently leveraging action structure information, our model can generate high-quality long-range action sequences with pure Gaussian noise and provided labels as inputs without pretraining. 
    \item Our model is evaluated on two standard large datasets for skeleton-based action generation, achieving superior and stable performance compared with previous models.
\end{itemize}





\section{Preliminaries \& Related Work}

\subsection{Attention Model}
Attention models have become increasingly popular in capturing long-term global dependencies \cite{bahdanau2014neural,chen2017pixelsnail}. In particular, self-attention \cite{brock2018large,vaswani2017attention,zhang2019self} mimics human visual attention, allowing a model to focus on crucial regions and to learn the correlation among elements in the same sequence. \cite{wang2018non} proposes a non-local operation as a kind of attention on capturing long-range dependencies in videos. \cite{vaswani2017attention} develops the transformer model, which is solely based on attention and achieves state-of-the-art on machine translation.  Thus, self-attention can typically lead to a better representation learning. One key advance of our proposed model compared with previous ones is that we adopt self attention to efficiently encode frame-wise correlations by inheriting all merits of the self-attention mechanism. 


\subsection{Skeleton-Based Action Generation}
The task of action generation differs from action prediction \cite{barsoum2018hp} in that no past intermediate sub-sequence is provided. Directly generating human actions from noise is considered more challenging. The problem has been well studied in early works \cite{bissacco2009hybrid,oh2005learning,pavlovic2001learning}, which applied switching linear models to generate stochastic human motions. These models, however, required a large amount of data to fit a model and are difficult to find an appropriate number of switching states. Later on, the Restricted Boltzmann Machine \cite{taylor2011two} and Gaussian-process latent variable models \cite{urtasun2008topologically,wang2009optimizing,wang2010optimizing} were applied. But they still can not scale to massive amounts of data. 
The rapid development of deep generative models has brought the idea of recurrent-neural-network (RNN) based Variational Autoencoder (VAE) \cite{kingma2013auto} and Generative Adversarial Net (GAN) models \cite{goodfellow2014generative,cai2018deep,kiasari2018human,wang2019combining,wang2019learning,wichers2018hierarchical,zhao2020feature}. These models are scalable and usually can generate actions with better quality. 

The aforementioned methods still have some limitations, which mainly lie in two aspects. Firstly, spatial relationships among body joints and temporal dynamics along continuous frames have not been well explored. 
Secondly, these models often require an expensive pre-training phase to capture intra-frame constraints, including the two most recent state-of-the-art works \cite{cai2018deep,wang2019learning}. By contrast, Our work moves beyond these limitations and can be trained from scratch to generate high-quality motions.

\begin{wrapfigure}{R}{0.5\linewidth}
    \centering 
    \includegraphics[width=\linewidth]{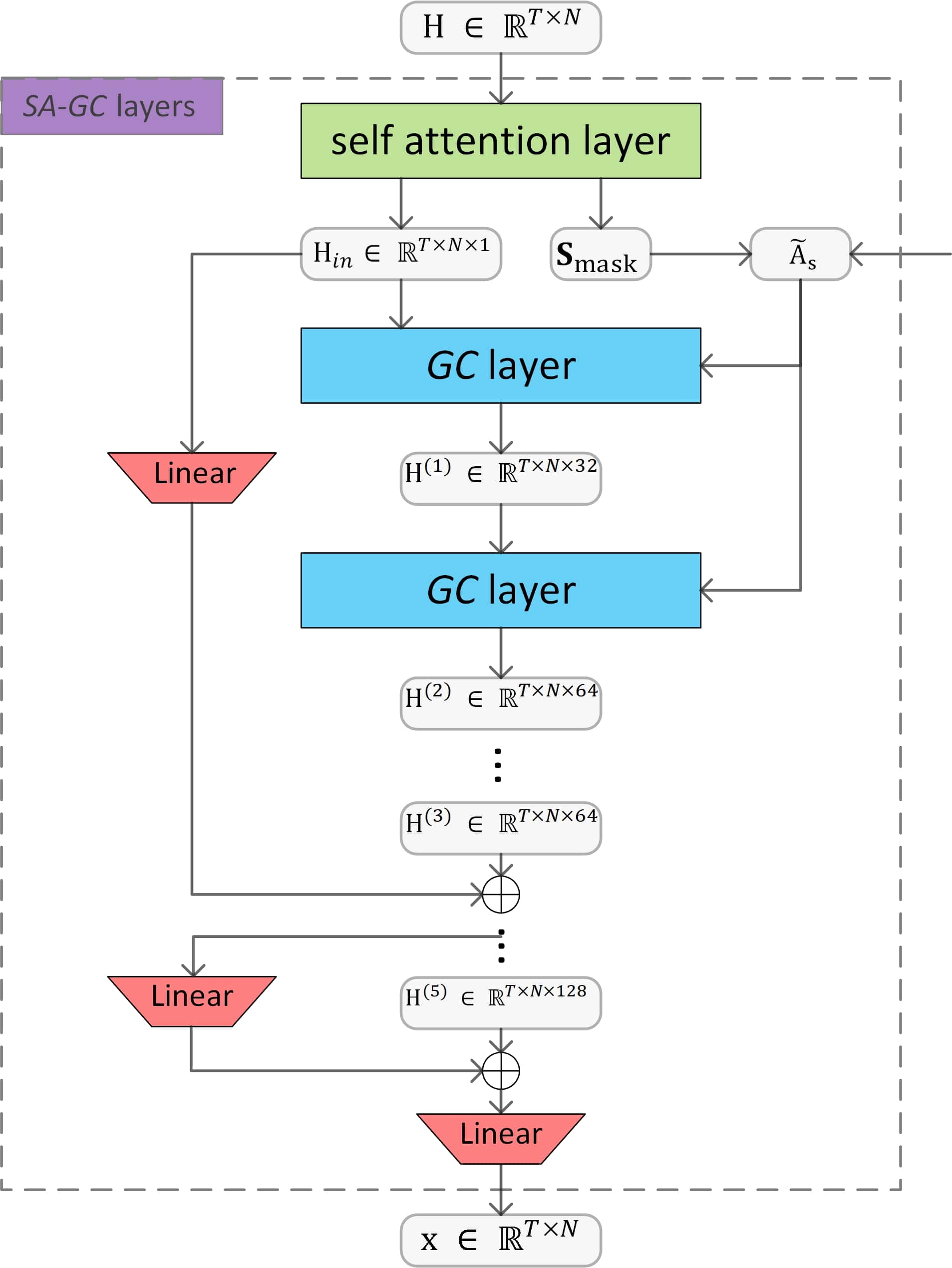} 
    \caption{\small{An illustration of the {\it SA-GC} layer. $\tilde{\mathbf{A}}$ and $\tilde{\mathbf{A}}_s$ are two adjacency matrices detailed in Section~\ref{sec:sa_gcn_layer}.}} 
    \label{fig:sa_gc} 
\end{wrapfigure}

\subsection{Graph Convolutional Network}
GCNs have been achieving encouraging results \cite{yan2018spatial}. In general, they can be categorized into two types: spectral approaches \cite{bruna2013spectral,kipf2016semi} and spatial approaches \cite{zhou2018graph,niepert2016learning}. The spectral GCN operates on the Fourier domain (locality) with convolution to produce a spectral representation of graphs. The spatial GCN, by contrast, directly applies convolution on the spatially distributed nodes. This work is in the spirit of spatial GCNs and incorporates new ideas of GCNs to fit the task. In particular, to model long-term dependent motion dynamics, we are aware of ideas from graph pruning \cite{zhang2018graph} and jump connection \cite{xu2018representation}, which respectively allows one to extract structure representation more efficiently and to build deeper graph convolutional layers. In terms of GCN-based human motion modelling, the most related work is {\it ST-GCN} \cite{yan2018spatial}, which applies a spatial GCN to a different task of action recognition. This method applied a GCN layer for intra-frame skeletons and then used 1D convolution layer for gathering information in temporal space. All nodes in a frame share weights on the temporal space and could only attend limited range of information, depending on the kernel size of the 1D convolution layer. We will compare our method with {\it ST-GCN} (for action generation) in Section~\ref{sec:ablation}.

\section{Structure-Aware Human-Action Generation}
Different from the video-generation task, the skeleton-based action generation contains huge amounts of structure information, {\it e.g.}, intra-frame structural joints information and inter-frame motion dynamics. Directly treating skeleton frames as images will lose most of these structure information, leading to the distortion of some skeletal frames. Moreover, in the context of skeleton-based actions, where only limited positional information is provided, differences between two continuous frames are virtually impossible to be observed.
To address these issues, we propose to incorporate GCNs to encode the rich structural information, with additional consideration to reduce computational burden by using self-attention to automatically learn a sparse action graph. 

\begin{figure}[!t]
    \centering
    \includegraphics[width=0.95\linewidth]{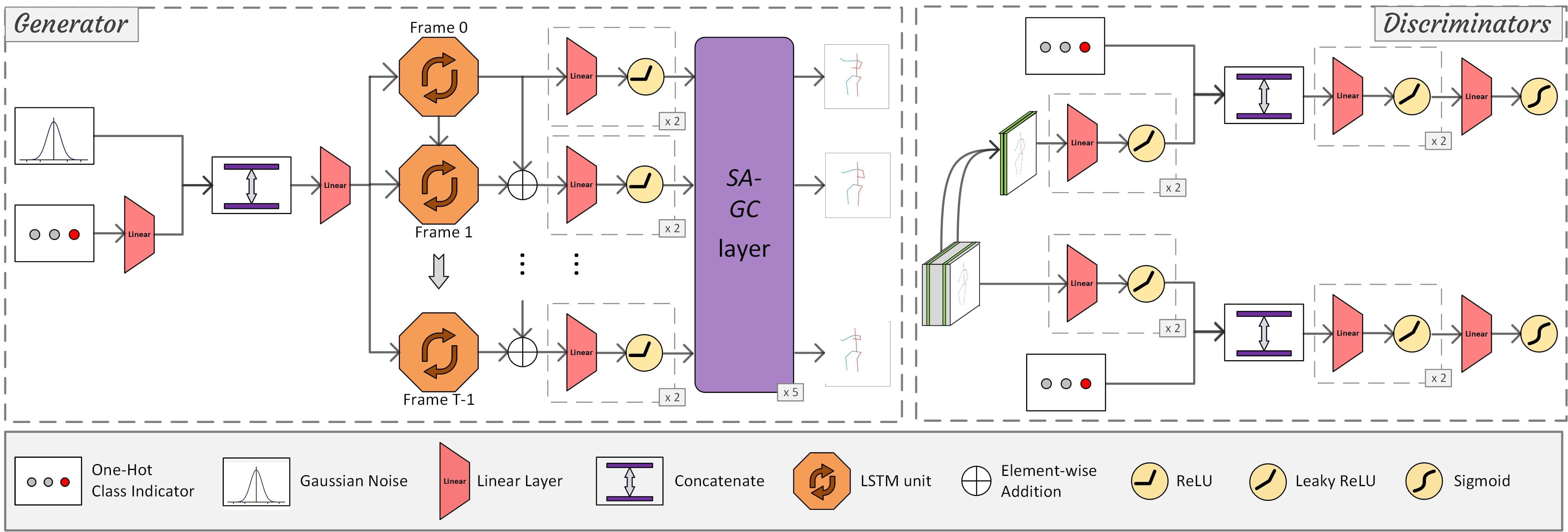}
    \caption{\small{The overall framework of the proposed method.}}
    \label{fig:framework}
\end{figure}

\subsection{An Overview of the {\it SA-GCN}}

Fig.~\ref{fig:framework} illustrates the overall framework of our model for action generation. It follows the GAN framework of video generation \cite{tulyakov2018mocogan,clark2019efficient}, which consists of an action generator  $\mathcal{G}$ and a dual discriminator: one video-based discriminator $\mathcal{D}_V$ and one frame-based discriminator $\mathcal{D}_F$.

\subsubsection{Generator}
For simplicity, we assume the sequence length to be $T$. 
Our action generator starts with a RNN with an input at each time as the concatenation of a Gaussian random noise $z$ and an embedded class representation of a label $y$. The outputs of the RNN layer are denoted as $[o_0, o_1, o_2, ... , o_{T-1}]$. Following \cite{cai2018deep,wang2019learning}, we consider outputting residuals instead of the exact coordinates of different joints, {\it i.e.}, $c_0 = o_0$, $c_1 = o_1 + c_0$, ..., $c_{T-1} = o_{T-1} + c_{T-2}$. The output of the RNN will go through three linear transformations before being fed as the input of the newly proposed {\it SA-GC} layer, which will be detailed in Section~\ref{sec:sa_gcn_layer}.

\subsubsection{{The \it SA-GC} layer}
The key component of our framework is a newly defined self-attention based graph convolutional layers ({\it SA-GC} layers), as illustrated in Fig.~\ref{fig:sa_gc}. Specifically, we denote the input of the {\it SA-GC} layers as a feature vector $\mathbf{H} \in \mathbb{R}^{T \times N}$. Through a self attention layer \cite{vaswani2017attention}, the output are a new representation $\mathbf{H}_{in} \in \mathbb{R}^{T \times N \times 1}$ and a learned masked attention score matrix 
$\mathbf{S}_{mask} \in \mathbb{R}^{T \times T}$. This self attention layer is followed by 5 {\it GC} layers. Each {\it GC} layer takes last layer's hidden state vector and masked adjacency matrix $\mathbf{\tilde{A}_s}$ as the input. The hidden states, which are outputs of the 5 {\it GC} layers, are defined respectively as $\mathbf{H}^{(1)} \in \mathbb{R}^{T \times N \times 32}$, $\mathbf{H}^{(2)} \in \mathbb{R}^{T \times N \times 64}$, $\mathbf{H}^{(3)} \in \mathbb{R}^{T \times N \times 64}$, $\mathbf{H}^{(4)} \in \mathbb{R}^{T \times N \times 128}$ and $\mathbf{H}^{(5)} \in \mathbb{R}^{T \times N \times 128}$.
Furthermore, the ResNet mechanism \cite{he2016deep} is applied on each two {\it SA-GC} layers, {\it i.e.}, we add the output of the first {\it SA-GC} layer to the third {\it SA-GC} layer, and the output of the third {\it SA-GC} layer to the final output. Detailed operations of the {\it SA-GC} layer are described in Section~\ref{sec:sa_gcn_layer}.

\subsubsection{Dual discriminator}
The video-based discriminator $\mathcal{D}_{V}$ takes a sequence of actions and the corresponding labels as the input. The frame-based discriminator $\mathcal{D}_{F}$ randomly selects $k_{frame}$ frames of an input sequence and the corresponding labels as the input. Both discriminators output either real or fake. 
In this paper, we apply the conditional GAN objective formulation \cite{goodfellow2014generative,mirza2014conditional,li2017alice}:

\begin{equation}
\begin{aligned}
\mathcal{L} = \min _{\mathcal{G}} \max _{\mathcal{D}_{F}, \mathcal{D}_{V}} &\mathbb{E}_{x \sim p(x)}[\log \mathcal{D}_{F}(x|y))]+\mathbb{E}_{z \sim p(z)}[\log (1-\mathcal{D}_{F}(\mathcal{G}(z|y)))]+\\
&\mathbb{E}_{x \sim p(x)}[\log \mathcal{D}_{V}(x|y)]+\mathbb{E}_{z \sim p(z)}[\log (1-\mathcal{D}_{V}(\mathcal{G}(z|y))))]
\end{aligned}
\end{equation}
where $p(x)$ defines the ground truth distribution, $p(z)$ is the standard Gaussian distribution and $y$ is the one-hot class indicator.



\begin{figure}[!b]
    \centering
    \includegraphics[width=0.95\linewidth]{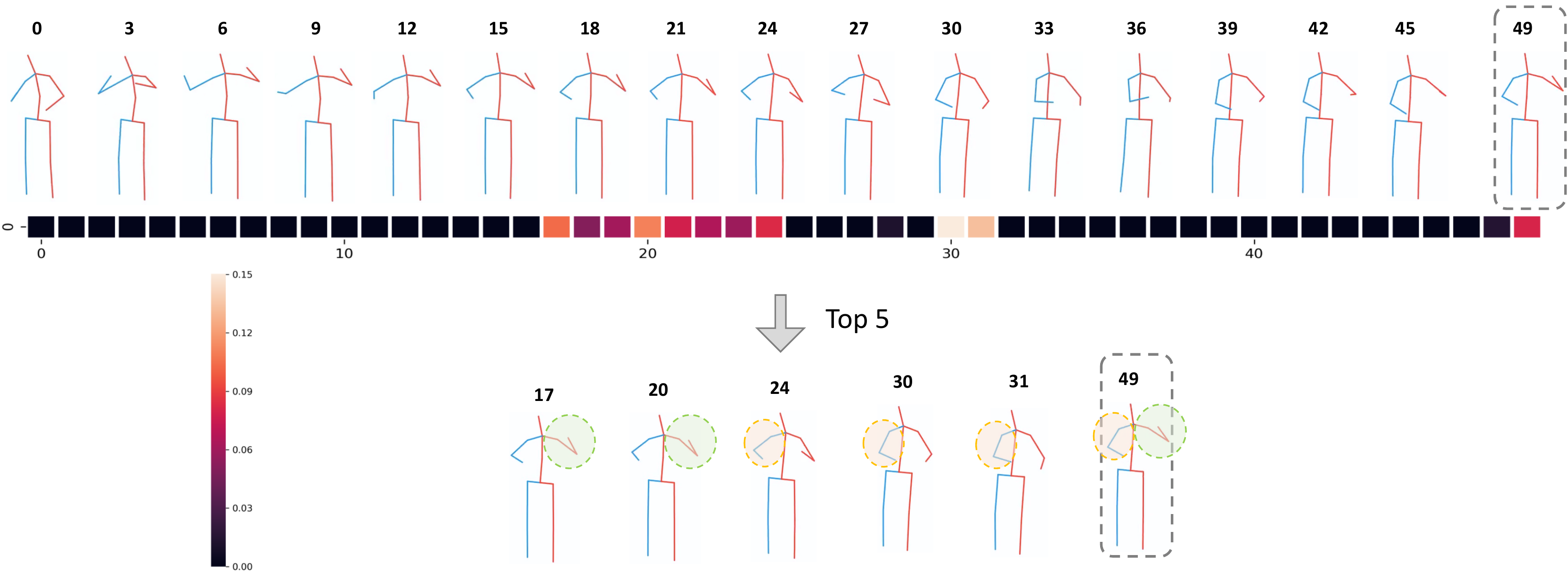}
    \caption{\small{The pipeline of {\it SA-GC} layer. The top line shows frames out of every three consecutive frames from Human 3.6 {\it Direction} class. The heat map under these samples represent the corresponding attention scores for the 49${th}$ frame. The bottom line shows the top 5 frames with the highest attention scores. Green circles and orange circles show similarity between selected frames and our target frame (the 49$th$ frame). }}
    \label{fig:frame_att}
\end{figure}

\subsection{Action Graph Construction}\label{sec:agc}
In this section, we describe detailed construction of the action graph, which is used in our {\it SA-GCN} module. Note a skeleton sequence is usually represented by 2D or 3D coordinates of human joints in each frame. The inter-frame action is the static skeleton in spatial space, and the inter-frame action is the movement in temporal space. 
To capture the temporal correlation, previous work has applied 1D convolution for learning skeleton representing by concatenating coordinate vectors of all joints in one frame. 
In our framework, as stated before, we propose to construct a connected graph for a whole action sequence, and learn a sparse inter-frame connection by adopting self-attention learning. Particularly, we construct an undirected graph \textbf{G=(V, E)} on a whole action sequence of $T$ frames, each consists of $N$ joints. Here, the node set $
V=\left\{v_{t i} | t=0, \ldots, T-1, i=1, \ldots N\right\}
$ includes all joints of a skeleton sequence. 

\subsubsection{Explanation of the {\it SA-GC} layer} \label{sec:sa_gcn_layer}

Fig.~\ref{fig:sa_gc} shows the detailed implementation of the {\it SA-GC} layer. Our {\it SA-GC} layer consists of one self attention layer and 5 graph convolution ({\it GC}) layers. To explain the construction, we detail the pipeline of the construction with an example illustrated in Fig.~\ref{fig:frame_att}.  

\paragraph{The Self attention layer}
Similar to standard self attention \cite{vaswani2017attention}, our self-attention layer takes a feature vector $\mathbf{H} \in \mathbb{R}^{T \times N}$ as input, and outputs a self-attention matrix $\mathbf{S}_{mask}$, representing how much influence of the past frames on the current frame. Fig.~\ref{fig:frame_att} shows one of our generated {\it Direction} sequences and its corresponding attention score vector's heat map for the last frame (the 49$th$ frame). After the self attention layer, we select top 5 past frames with the highest attention scores (only keep 6 elements in each row of the $\mathbf{S}_{mask}$ matrix). As we could see from the example in Fig.~\ref{fig:frame_att}, the selected 5 past frames have the highest similarity with the 49$th$ frame. The skeleton in the 49${th}$ frame keeps the red arm up and keeps the blue arm bent down. Looking back to past frames, frames before the 21${st}$ lift up its red arm. Frames between the 24${th}$ to the 31${st}$ frame have the similar blue arm pose as the 49${th}$ frame. Our attention identifies frames 17${th}$, 20${th}$, 24${th}$, 30${th}$ and 31${st}$ as the most relevant frames to be attended according to the learned attention matrix $\mathbf{S}_{mask}$.
\begin{figure}[!t]
    \centering
    \includegraphics[width=0.85\linewidth]{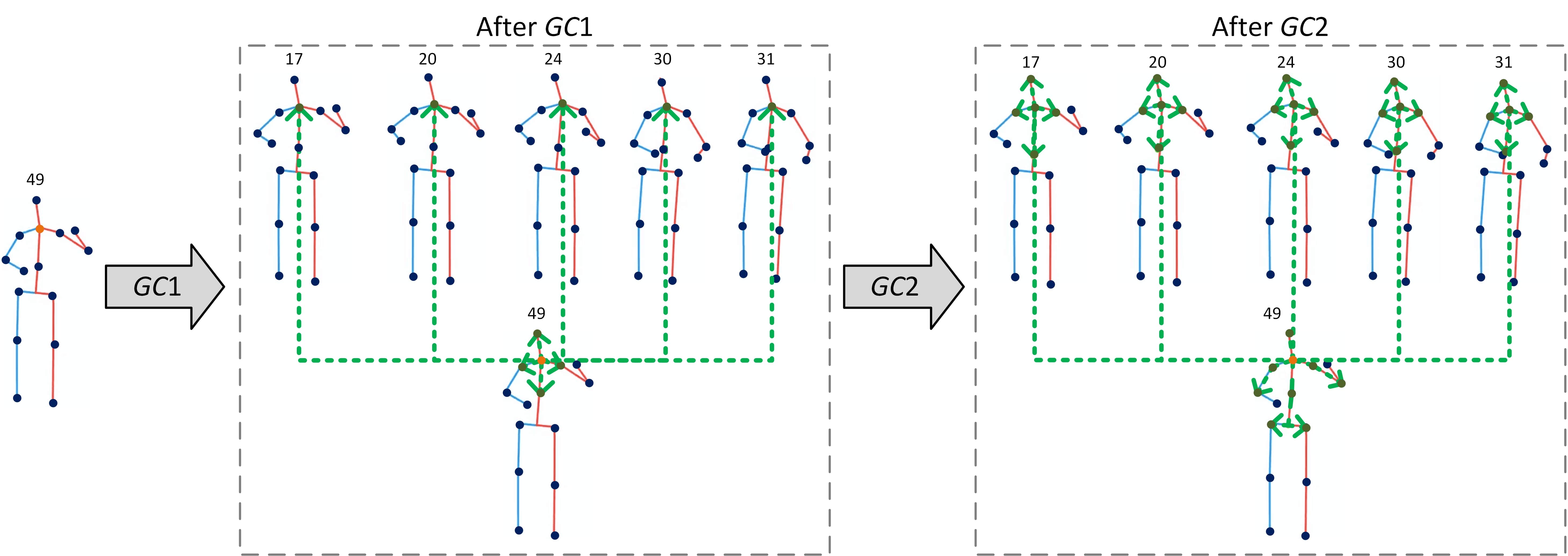}
    \caption{\small{Information passing through {\it SA-GC} layers at the node ``neck''.}}
    \label{fig:gc_att}
\end{figure}
\paragraph{The {\it GC} layers}
As illustrated, the self-attention layer is followed by 5 {\it GC} layers. 
After selection, we will connect each node of the 49${th}$ frame with the corresponding node in the selected 5 frames and assign edge weights with the corresponding self-attention scores. Fig.~~\ref{fig:gc_att} shows information passing path through our {\it SA-GC} layer at the node neck. The left plot of Fig.~\ref{fig:gc_att} shows that after one {\it GC} layer, the 49${th}$ neck node can gather information from neck nodes of five selected frames and four neighbor nodes in its own frame. The right plot of Fig.~\ref{fig:gc_att} shows that after the two {\it SA-GC} layers, the 49${th}$ neck node can gather information for the five nodes of the selected past four frames and seven nodes of its own frame. 
It is worth noting that nodes in different frame will have distinct attention score for edges in both spatial space and temporal space, thus they will have their particular edge weights through our {\it SA-GC} layer. 

\subsubsection{Implementing self-attention based GCN}
In our case, we consider all joints in an action sequence, ending up with a 2D adjacent matrix with both row size and column size $N*T$. 
To this end, we first use $\mathbf{A} \in \mathbb{R}^{N \times N}$ to denote the adjacent matrix of intra-frame, which is constructed by strictly following the structure of a skeleton, {\it e.g.}, the ``head'' node is connected to the ``neck'' node. 
After adding self connections $\mathbf{I}$, the intra-frame adjacency matrix will be $\mathbf{\bar{A}} = \mathbf{A}+ \mathbf{I}$.
We then define an initial adjacency matrix of a whole sequence as: 

\begin{equation}
\tilde{\mathbf{A}} = \left(\begin{array}{ccccc}
{\bf \bar{A}} & {\bf I} & \cdots & {\bf I} \\
{\bf I} & {\bf \bar{A}} & \cdots & {\bf I} \\
\vdots & \vdots & & \vdots \\
{\bf I} & {\bf I} & \cdots & {\bf \bar{A}}
\end{array}\right)_{(N*T) \times (N*T)} ,
\end{equation}
where $\mathbf{I}$ is used to represent connecting each node with all of the corresponding nodes in the temporal space, $(N*T) \times (N*T)$ means $\mathbf{\tilde{A}}$ is a 2D matrix with both row size and column size $N*T$, both $N$ and $T$ are numbers, * means multiply operation. The adjacency matrix $\mathbf{\tilde{A}}$ essentially means each node in one frame is connected to the corresponding node in the temporal space. 
At the same time, it also connects to the neighboring nodes in spatial space encoded by $\mathbf{\bar{A}}$.  

Next, we propose to use self-attention to prune the action graph. The idea is to learn a set of attention scores encoding the relevance of each frame w.r.t.\! the current frame, and only choose the top-$K$ frames in the temporal space. Specifically, we adopt a similar implementation of the scaled dot-product attention as in \cite{vaswani2017attention}. The input of the self-attention layer is represented as $\mathbf{H} \triangleq \{h_0, h_1, \cdots, h_{T-1}\}$, where $h_t \in \mathbb{R}^{N}$ represents the hidden state vector at time $t$ with $N$ nodes.
Following the self-attention in \cite{vaswani2017attention}, $\mathbf{Q}$, $\mathbf{K}$ and $\mathbf{V}$ are given as:
\begin{equation}
    \mathbf{Q} = \mathbf{W}_q\mathbf{H} \; , \;
    \mathbf{K} = \mathbf{W}_k\mathbf{H} \; , \;
    \mathbf{V} = \mathbf{W}_v\mathbf{H} \; , 
\end{equation}
where $\mathbf{W}_q$, $\mathbf{W}_k$ and $\mathbf{W}_v$ are projection weights. The attention score $\mathbf{S} \in \mathbb{R}^{T \times T}$ and the attention layer's output $\mathbf{H}_{in}$ are calculated as:
\begin{equation}
\mathbf{S} = \operatorname{softmax}\left(\mathbf{Q} \mathbf{K}^\mathsf{T}\right); ~\mathbf{H}_{in}=\mathbf{SV}
\end{equation}

In the task of action generation, we need to modify $\mathbf{S}$ as a masked attention $\mathbf{S}_\textit{mask}$ which prevents current frame from attending to subsequent frames
\begin{equation}
\begin{small}
\mathbf{S}_{mask} = \left(\begin{array}{ccccc}
s_{0,0} & 0 & \cdots & 0 \\
s_{1,0} & s_{1,1} & \cdots & 0 \\
\vdots & \vdots & & \vdots \\
s_{T-1,0} & s_{T-1,1} & \cdots & s_{T-1,T-1}
\end{array}\right)_{T \times T} ,
\label{eq:mask_attention_score}
\end{small}
\end{equation}
where the element $s_{m,n}$ denotes the {\it n}-th frame's influence on the {\it m}-th frame and values in the upper triangle are all equal to 0.
To enforce the pruning, we further select the top $K$ scores in each row of the $\mathbf{S}_{mask}$ and set the other elements to be 0. 
Note that, if the number of non-zero elements in some rows is less than {\it K}, we will keep all the non-zero elements. Finally, the adjacent matrix is constructed as
\begin{equation}
\begin{small}
\mathbf{\tilde{A}}_s = \mathbf{S}_{mask}\odot\tilde{\mathbf{A}} \triangleq \left(\begin{array}{ccccc}
s_{0,0}*\mathbf{\bar{A}} & \mathbf{0} & \cdots & \mathbf{0} \\
s_{1,0}*\mathbf{I}& s_{1,1}*\mathbf{\bar{A}} & \cdots & \mathbf{0} \\
\vdots & \vdots & & \vdots \\
s_{T-1,0}*\mathbf{I} & s_{T-1,1}*\mathbf{I} & \cdots & s_{T-1,T-1}*\mathbf{\bar{A}}
\end{array}\right)_{(N*T) \times (N*T)}
\end{small}
\end{equation}
Consequently, the output (before activation) of the self-attention based graph convolutional layer becomes:
\begin{equation}
\begin{small}
\mathbf{H}^{(1)}=\mathbf{D}^{-1} \mathbf{\tilde{A}}_s \mathbf{H}_{in}\mathbf{W} \, ,
\label{eq:gcn}
\end{small}
\end{equation}
where $\mathbf{D}^{i i}=\sum_{j} \mathbf{\tilde{A}}_s^{i j}$ represents diagonal node degree matrix for normalizing $\mathbf{\tilde{A}}_s$, $\mathbf{H}^{(1)}$ is the hidden state after first \textbf{GC}
layer in Fig.~\ref{fig:sa_gc}. 
We will conduct graph convolution operation in equation \ref{eq:gcn} using same $\mathbf{\tilde{A}}_s$ for five times. The output of the fifth {\it GC} layer in {\it SA-GC} layer is $\mathbf{H}^{(5)}$. After a linear layer, we get the output of the generator, which is a generated sequence $x \in \mathbb{R}^{T \times N}$.


\section{Experiments}
We perform experiments to evaluate the proposed method on two standard skeleton-based human-action benchmarks, the Human-3.6m dataset \cite{ionescu2013human3} and the NTU RGB+D dataset \cite{shahroudy2016ntu}. Several state-of-the-art methods are used for comparison, including \cite{wichers2018hierarchical,habibie2017recurrent,cai2018deep,wang2019learning}. Following \cite{wang2019learning}, the Maximum Mean Discrepancy (MMD) \cite{gretton2012kernel} is adopted to measure the quality of generated actions. Further, we pre-train a classifier on training set to test the recognition accuracy of generated actions.
We also conduct human evaluation on the Amazon Mechanical Turk (AMT) to access the perceptual quality of generated sequences. To examine the functionality of each component of the proposed model, we also perform detailed ablation studies on the Human-3.6m dataset.

\subsection{Datasets}
\paragraph{Human-3.6m} 
Following the same pre-processing procedure in \cite{cai2018deep,wang2019learning}, 50 Hz video frames are down-sampled to 16 Hz to obtain representative and larger variation 2D human motions. The joint information consists of 2-D locations of 15 major body joints. Ten distinctive classes of actions are used in the following experiments, including \textit{sitting, sitting down, discussion, walking, greeting, direction, phoning, eating, smoking} and \textit{posing}.

\paragraph{NTU RGB+D}
This dataset contains 56,000 video clips on 60 classes performed by 40 subjects and recorded with 3 different camera views. Compared with Human-3.6m, it can provide more samples in each class and much more intra-class variations. We select ten classes of motions and obtain their 2-D coordinates of 25 body joints following the same setting in \cite{wang2019learning}, including \textit{drinking water, jump up, make phone call, hand waving, standing up, wear jacket, sitting down, throw, cross hand in front} and \textit{kicking something}. We then apply two commonly used benchmarks for a further evaluation in the ten classes:
$(\RN{1})$\textit{cross-view}: the training set contains actions captured by two cameras and remaining data are left for testing.
$(\RN{2})$\textit{cross-subject}: action clips performed by 20 subjects are randomly picked for training and another 20 subjects are reserved for testing.

\subsection{Training Details}

Following \cite{wang2019learning}, we set the action sequence length for both datasets to be 50. The image discriminator randomly selects 20 frames from every generated sequence and training sequences as the input. The {\it SA-GC} layer selects top 5 past frames to construct an adjacency matrix $\mathbf{\tilde{A}_s}$. We set batch size for training to be 100, for testing to be 1000, and the learning rate to be 0.0002. 

\subsection{Evaluation Metrics}
\paragraph{Maximum Mean Discrepancy} 
The MMD metric is based on a two-sample test to measure the similarity between two distributions $\mathcal{P}(x)$ and $\mathcal{Q}(y)$, based on samples $x\sim \mathcal{P}(x)~\text{and}~y \sim \mathcal{Q}(y)$. It is widely used to measure the quality of generated samples compared with real data in deep generative model \cite{zhao2019self} and Bayesian sampling \cite{han2018stein}. The metric has also been applied to evaluate the similarity between generated actions and the ground truth in \cite{walker2017pose,wang2019learning}, which has been proved consistent with human evaluation. As motion dynamics are in the form of sequential data points, we denote $\text{MMD}_{\text{avg}}$ as the average MMD over each frame and $\text{MMD}_{\text{seq}}$ to denote the MMD over whole sequences.

\paragraph{Recognition Accuracy}
Apart from using MMD to evaluate the model performance, we also pre-train a recognition network on the training data to compute the classification accuracy of generated samples. The recognition network exactly follows the video discriminator except for the last \textit{softmax} layer. This evaluation metrics can examine whether the conditional generated samples are actually residing in the same manifold as the ground truth and can be correctly recognized. Details are given in the Appendix.

\subsection{Baselines}
We compare our method with six baselines. We first consider the model in \cite{wichers2018hierarchical}, which can be used to generate long-term skeleton-based actions in an end-to-end manner. This includes three training alternatives: end-to-end ({\it E2E}), E2E prediction with visual analogy network ({\it EPVA}) and EPVA with adversarial loss ({\it adv-EPVA}). The second baseline \cite{habibie2017recurrent} is based on VAE, called the {\it SkeletonVAE}, which improves previous motion generation methods significantly. Finally, two most recent strong baselines are considered, including the previous state-of-the-art method \cite{cai2018deep} and an improved version \cite{wang2019learning} with an auxiliary classifier. The latter utilizes a \textit{Single Pose Training} stage and a \textit{Pose Sequence Generation} stage to produce high-quality motions. These two baselines are respectively referred to as {\it SkeletonGAN} and {\it c-SkeletonGAN}.

\subsection{Detailed Results}
\subsubsection{Quantitative results}
Our {\it SA-GCN} model shows superior quantitative results in terms of both MMD and recognition accuracies on the two datasets, compared with related baseline models.
\paragraph{Human-3.6m}
Table~\ref{tab:mmd-hm36} shows MMD results of our model and the baselines on Human-3.6m. With structure information considered, our model achieves significant performance gains over all baselines, which even without the need of an inefficient pre-training stage. The recognition accuracies are reported in Table~\ref{tab:acc_hm36}. Similarly, our model consistently outperforms three baselines by a large margin. Please note none information of the generated actions are used in the pretrained classifier, thus we avocate that the relatively low recognition accuracies are indeed reasonable. On the other hand, this also indicates that existing action generation models are still far from satisfactory.
\paragraph{NTU RGB+D}
This dataset is more challenging, which contains more body joints and action variations. In the experiments, we find that three models (E2E, EPVA, adv-EPVA \cite{wichers2018hierarchical}) fail to generate any interpretable action sequences. As a result, we only present MMD results for the other three baselines in Table~\ref{tab:mmd-ntu}. Again, the proposed method performs the best among all models under \textit{cross-view} and \textit{cross-subject} settings. 
\begin{table}[!t]
    \centering
    \caption{\small{Model comparisons in terms of MMD on Human-3.6m.}}
    \scalebox{0.75}{
    \begin{tabular}{c|c|c|c}
    \toprule
        Models & Pretrain & $\text{MMD}_{\text{avg}}\downarrow$ & $\text{MMD}_{\text{seq}}\downarrow$ \\ \midrule
        \textit{E2E} \cite{wichers2018hierarchical} & No & 0.991 & 0.805 \\ 
        \textit{EPVA} \cite{wichers2018hierarchical} & No & 0.996 & 0.806 \\ 
        \textit{adv-EPVA} \cite{wichers2018hierarchical} & No& 0.977 & 0.792 \\ 
        \textit{SkeletonVAE} \cite{habibie2017recurrent}  & No & 0.452 & 0.467 \\ 
        \textit{SkeletonGAN} \cite{cai2018deep} & Yes & 0.419 & 0.436 \\
        \textit{c-SkeletonGAN} \cite{wang2019learning} & Yes & 0.195 & 0.218 \\ \midrule
        Ours & No & \textcolor{tblue}{\bf{0.146}}  &  \textcolor{tblue}{\bf{0.134}} \\ \bottomrule 
    \end{tabular}
    }
    \label{tab:mmd-hm36}
\end{table}
\begin{table*}[!t]
    \centering
    \caption{\small{Action recognition accuracy on the generated actions on the Human-3.6m.}}
    \scalebox{0.75}{
    \begin{tabular}{c|cccccccccc|c}
    \toprule
        Models & Direct & Discuss & Eat & Greet & Phone & Pose & Sit & SitD & Smoke & Walk &  Average \\ \midrule
        \textit{SkeletonVAE} & 0.37 & 0.01 & 0.51 & 0.47 & 0.10 & 0.03 & 0.17 & 0.33 & 0.01 & 0.01 & 0.201 \\
        \textit{SkeletonGAN} & 0.35 & 0.29 & 0.72 & 0.66 & 0.46 & 0.09 & 0.32 & 0.71 & 0.14 & 0.02 & 0.376 \\
        \textit{c-SkeletonGAN}  & 0.34 & 0.44 & 0.57 & 0.56 & 0.52  & 0.25 & 0.67 & 1.00 & 0.50 &0.03 & 0.488 \\  \midrule
        \textit{SA-GCN} & 0.42 & 0.40 &0.78 & 0.55 & 0.72 & 0.61 & 0.95 & 0.79 & 0.52 & 0.18 & \textcolor{tblue}{{\bf 0.593}} \\
        \bottomrule
    \end{tabular}}
    \label{tab:acc_hm36}
\end{table*}
\begin{table}[!t]
    \centering
    \caption{\small{Model comparisons in terms of MMD on NTU RGB+D.}}
    \scalebox{0.75}{
    \begin{tabular}{c|c|c|c|c}
    \toprule
     \multirow{2}{*}{Models}& \multicolumn{2}{c}{\textit{cross-view}} & \multicolumn{2}{|c}{\textit{cross-subject}}\\ \cmidrule{2-5} 
        &$\text{MMD}_{\text{avg}}\downarrow$ & $\text{MMD}_{\text{seq}}\downarrow$ & $\text{MMD}_{\text{avg}}\downarrow$ & $\text{MMD}_{\text{seq}}\downarrow$ \\ \midrule
        \textit{SkeletonVAE} \cite{habibie2017recurrent}   & 1.079 & 1.205 & 0.992&1.136 \\ 
        \textit{SkeletonGAN} \cite{cai2018deep}  & 0.999 & 1.311  & 0.698 & 0.788  \\
        \textit{c-SkeletonGAN} \cite{wang2019learning}  & 0.371 & 0.398 & 0.338 & 0.402 \\ \midrule
        \textit{SA-GCN} &  \textcolor{tblue}{\bf{0.316}}  &  \textcolor{tblue}{\bf{0.335}} & \textcolor{tblue}{\bf{0.285}} & \textcolor{tblue}{\bf{0.299}} \\ \bottomrule 
    \end{tabular}
    }
    \label{tab:mmd-ntu}
\end{table}
        
\begin{figure}[!htbp]
    \centering
    \includegraphics[width=0.7\textwidth]{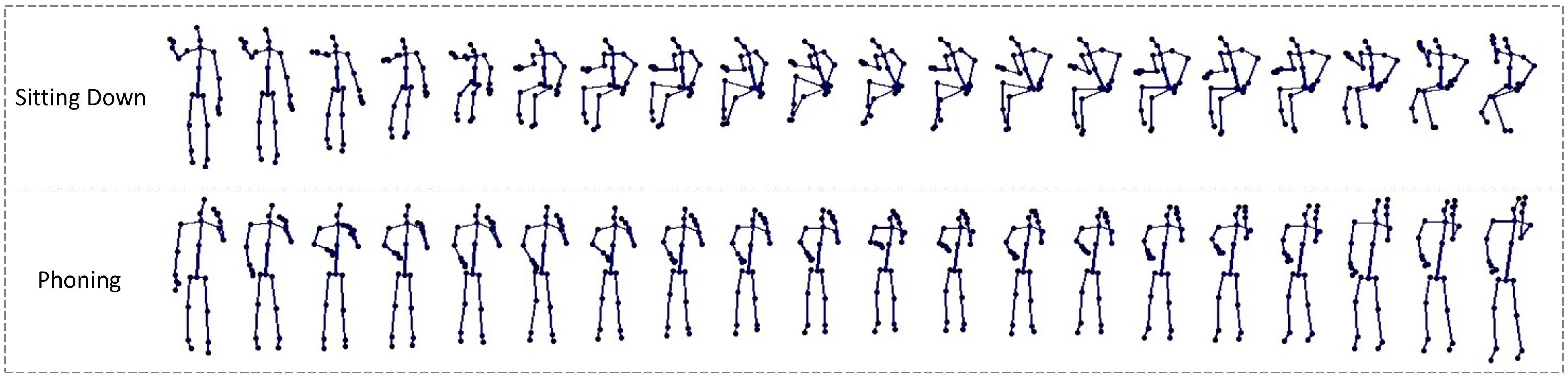}
    \caption{\small{Randomly selected samples on NTU RGB+D dataset. Top: \textit{sitting down} from \textit{cross-subject}, Bottom: \textit{phoning} from \textit{cross-view}.}}
    \label{fig:ntu-samples}
\end{figure}
\subsubsection{Qualitative results}
\begin{figure}[!htbp]
    \centering
    \includegraphics[width=0.75\textwidth]{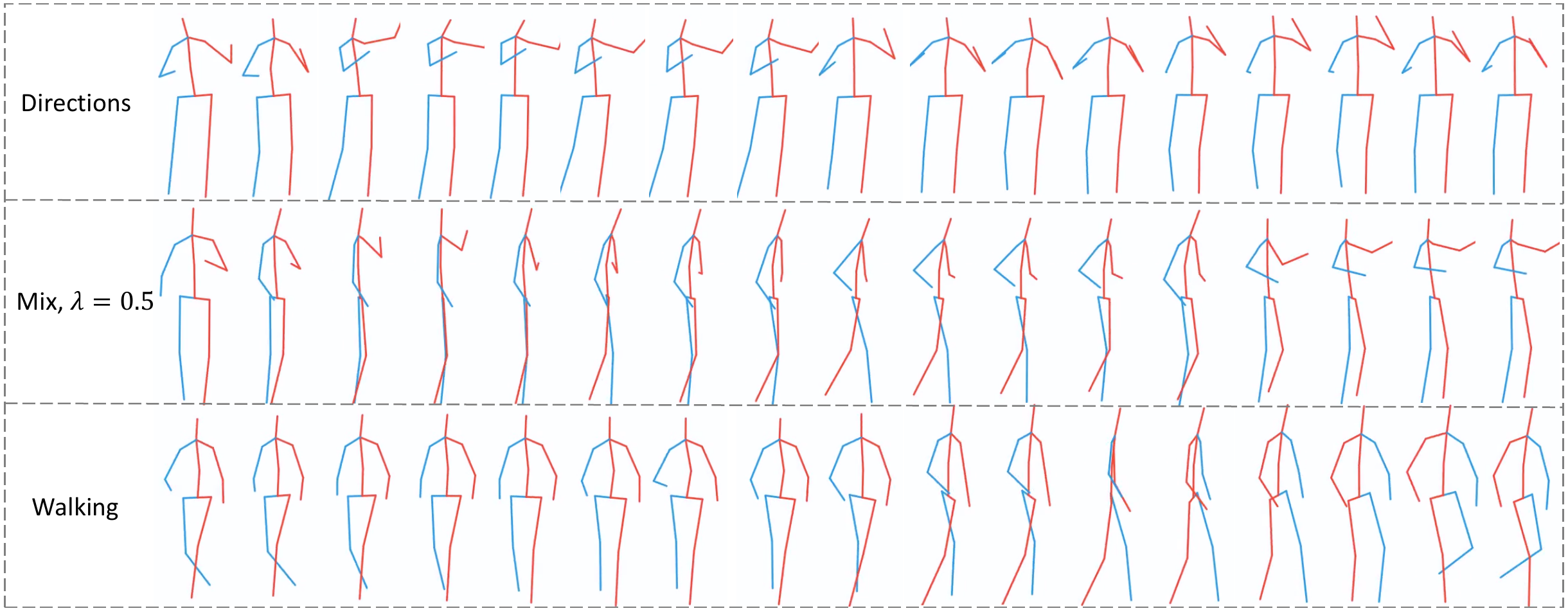}
    \caption{\small{Generated sequences of \textit{directions}, \textit{walking} and a mixed action with $\lambda=0.5$.}}
    \label{fig:mix-samples}
\end{figure}

We present some generated actions in Human-3.6m dataset and NTU RGB+D dateset in Fig.~\ref{fig:mix-samples}(first and third row) and Fig.~\ref{fig:ntu-samples} respectively. It is easy to see that our model can generate very realistic and easily recognizable actions. We also plot action trajectories on a projected space by t-SNE \cite{maaten2008visualizing} for each generated action class on the Human-3.6m dataset in Fig.~\ref{fig:traject}. It is observed that a group of actions, {\it i.e.}, \textit{directions, discussion}, \textit{greeting}, are close to each other, and so is the group \textit{sitting, sitting down}, \textit{eating}; while actions \textit{smoking} and \textit{sitting down} are far away. These are consistent with what we have observed in the ground truth.
\subsubsection{Smooth action generation}
\begin{figure}[!t]
    \begin{minipage}{.4\textwidth}
        \centering
        \includegraphics[width=\textwidth]{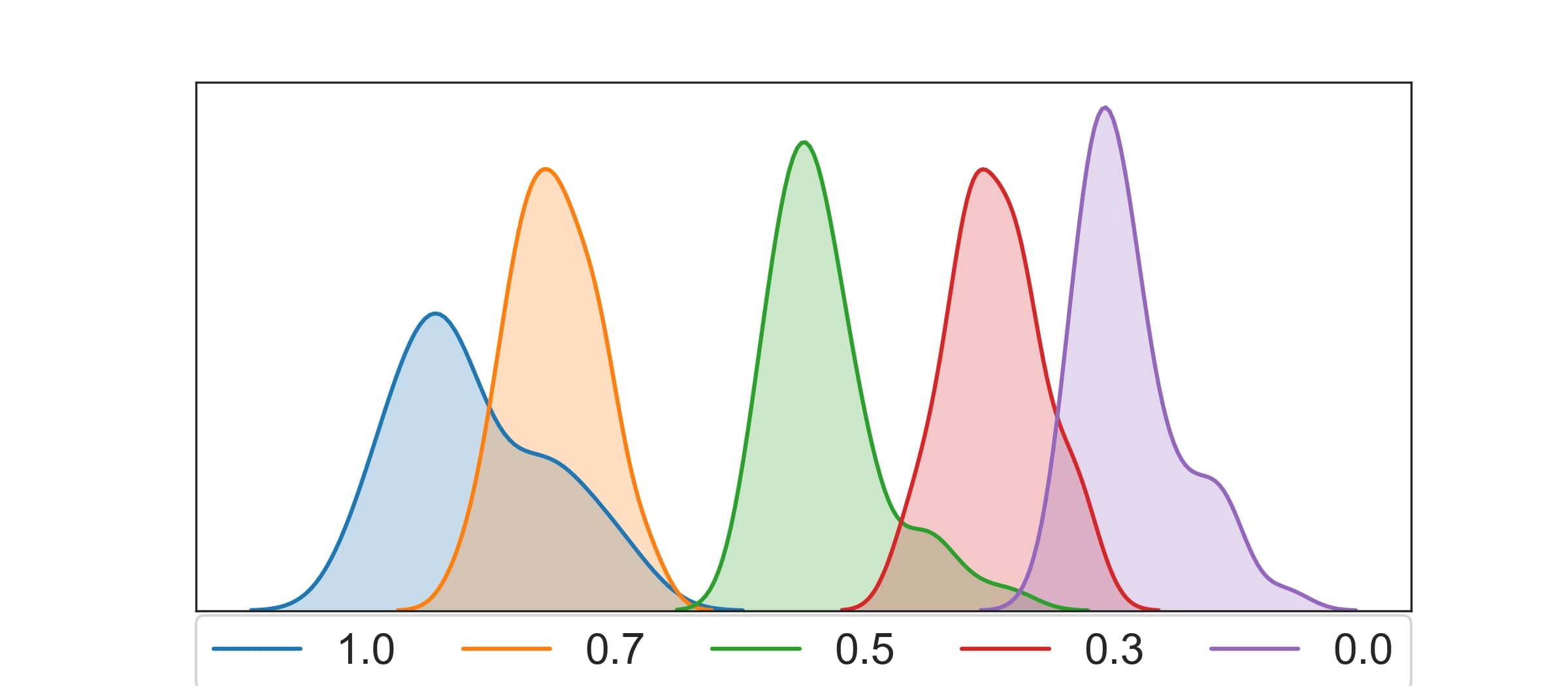}
        \captionof{figure}{\small Histogram of mixed actions where each mode represents an action with a smoothing term $\lambda$.
        } 
        \label{fig:mix}
    \end{minipage}%
    \hfill
    \begin{minipage}{.4\textwidth}
        \centering
        \includegraphics[width=0.95\textwidth]{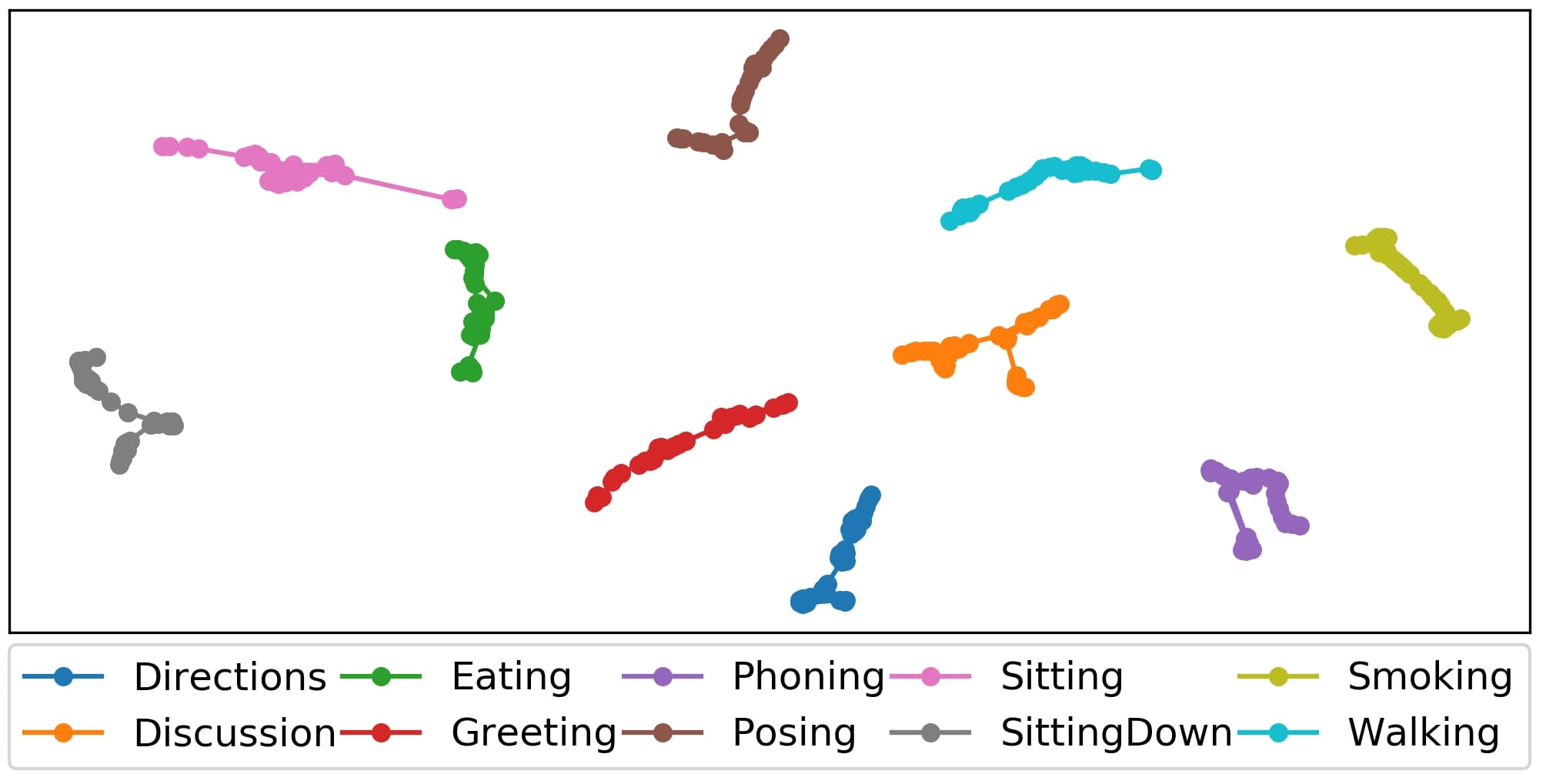}
        \captionof{figure}{\small Action trajectories on Human-3.6m.}
        \label{fig:traject}
    \end{minipage}
\end{figure}

Humans are capable of switching two actions very smoothly and naturally. 
For instance, a person can show others directions and walking at the same time. 
In this part, we verify that our model is expressive enough to perform such transitions as humans do. We use \eqref{eq:mix} to produce a smooth action transition between action classes $y_1$ and $y_2$ with a smoothing parameter $\lambda \in [0, 1]$. 
We generate 100 video clips with every mix and apply t-SNE \cite{maaten2008visualizing} to project the averaged sequences to a 1D manifold. The histogram of various mixed actions is shown in Fig.~\ref{fig:mix}. As we decrease $\lambda$, the mode (action) gradually moves from \textit{directions} towards \textit{walking}, meaning that our model can produce very smooth transitions when interpolating between the two actions. Fig.~\ref{fig:mix-samples} illustrates as randomly selected samples.
\begin{equation}\label{eq:mix}
    y_{mix} = \lambda y_1 + (1 - \lambda)y_2; \quad
    x_{mix} = G(z; y_{mix}), ~z \sim \mathcal{N}(0,1)
\end{equation}

\subsection{Ablation Study} \label{sec:ablation}
Our key innovation in our model is the {\it SA-GC} layer. As a result, we conduct detailed experiments to verify the effectiveness and usefulness of our self-attention based graph convolutional layer on Human 3.6m dataset. Since the self-attention layer has already been proved to be effective for sequential data, we keep the self-attention layer for all the following baselines. Without special mentioning, we keep all the other parts of the model to be the same.

\paragraph{Baseline 1: replace GCN layers with CNN layers}
We replace 5 GCN layers with 5 CNN layers using the same hidden dimension and kernel size. 

\paragraph{Baseline 2: without the inter-frame A matrix}
Based on our model, we drop the attention connections to past frames. That setting is the same as setting our top \textbf{k} to be 0 in our {\it SA-GC} layer. Under this Baseline, each frame in the sequence will be an independent graph for graph convolutional layer.  

\paragraph{Baseline 3: replace self-attention based GCN layers with the {\it ST-GC} layers \cite{yan2018spatial}}

The {\it ST-GC} layer leverage graph convolution for skeleton-based action recognition. Each {\it ST-GC} layer combines one graph convolution layer for learning intra-frame skeleton and one 1D convolutional layer for feature aggregation in the temporal space. 
%

The {\it Fully Connected} model described in Fig.~\ref{fig:introduction} is not applicable and can not scale to long sequences because it demands excessive amount of memory and computational resources.
The results of above three baselines are shown in Table~\ref{tab:ablation}. Comparing with baseline2 and baseline3, we can see that adding the adjacency matrix makes the model harder to train compared with CNN. However, our proposed self-attention can mitigate the difficulties and surpass standard CNN method on the skeleton based action generation task with much lower MMD scores.

\begin{table}[!h]
    \begin{minipage}{.55\textwidth}
    \centering
    \caption{\small{Ablation study results.}}
    \scalebox{0.95}{
    \begin{tabular}{c|c|c}
    \toprule
        Baselines & $\text{MMD}_{\text{avg}}\downarrow$ & $\text{MMD}_{\text{seq}}\downarrow$ \\ \midrule
        Baseline 1  & 0.240 & 0.222\\ 
        Baseline 2  & 0.915 & 0.922 \\ 
        Baseline 3  & 0.580 & 0.595 \\  \midrule
        Ours & \textcolor{tblue}{\bf{0.152}}  &  \textcolor{tblue}{\bf{0.142}}  \\ \bottomrule
    \end{tabular}
    }
    \label{tab:ablation}
    \end{minipage}%
    \hfill
    \begin{minipage}{.45\textwidth}
    \centering
    \caption{\small{AMT Evaluations}}
    \scalebox{0.95}{
    \begin{tabular}{c|c}
    \toprule 
        Models &  Evaluation Score$\uparrow$\\ \midrule
        \textit{SkeletonVAE} &  2.401\\ 
        \textit{SkeletonGAN} & 2.731 \\ 
        \textit{c-SkeletonGAN} &  3.157\\ \midrule 
        \textit{SA-GCN} &   \textcolor{tblue}{{\bf 3.925}} \\ 
    \bottomrule
    \end{tabular}
    }
    \label{tab:amt}
    \end{minipage}
\end{table}

\subsection{Human Evaluation}
We finally conduct perceptual human evaluations in the AMT platform. Four models are trained on the Human-3.6m dataset, including {\it SkeletonVAE}, {\it Skeleton-GAN}, {\it c-SkeletonGAN} and our {\it SA-GCN}. We then sample 100 action clips for each of the 10 action classes; 140 workers were asked to evaluate the quality of the generated sequences and score them in a range from 1 to 5. A higher score indicates a more realistic action clip. We only inform them of the action class and one real action clip to ensure proper judgements. The design detail is given in the Appendix. Table~\ref{tab:amt} demonstrates that our model is significantly better than other baselines in human evaluation. 

\section{Conclusions}

In this paper, we have presented the self-attention graph convolutional network ({\it SA-GCN}) to efficiently encode structure information into skeleton-based human action generation. Self-attention can capture long-range dependencies in continuous action sequences and learn to prune the dense action graph for efficient training. Further, the graph convolution is applied to seamlessly encode both spatial joints information and temporal dynamics information into the model. Based on these ideas, our model directly transforms noises to high-quality action sequences and can be trained end-to-end. On two standard human action datasets, we observe a significant improvement of generation quality in terms of both quantitative and qualitative evaluations. 



\clearpage
%
%
\bibliographystyle{splncs04}
\bibliography{main}

\newpage
\appendix

\section{Experiments Results}

We further show some action samples on both the Human-3.6m dataset \cite{ionescu2013human3} and the NTU RGB+D dataset \cite{shahroudy2016ntu}. We sample one frame in terms of every two consecutive frames to show the whole sequence of actions.

\begin{figure}[h!]
  \centering
  \begin{subfigure}{0.99\textwidth}
    \includegraphics[width=\textwidth]{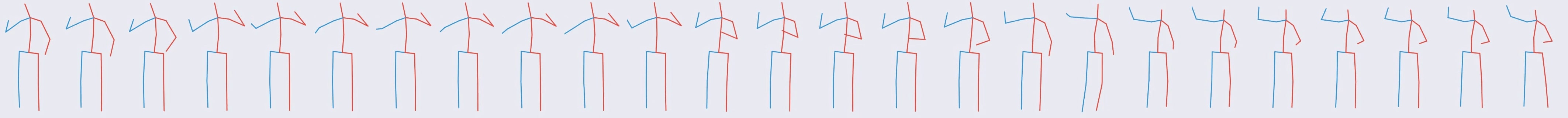}
  \end{subfigure}
  \begin{subfigure}{0.99\textwidth}
    \includegraphics[width=\textwidth]{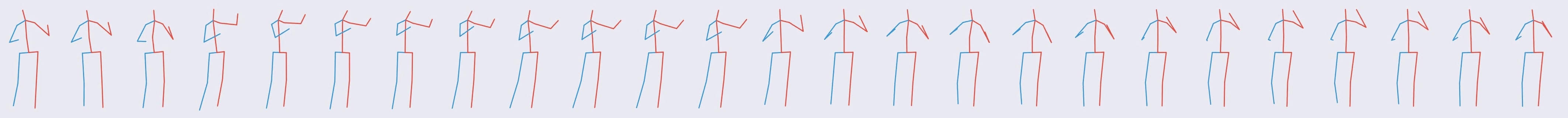}
  \end{subfigure}
  \begin{subfigure}{0.99\textwidth}
    \includegraphics[width=\textwidth]{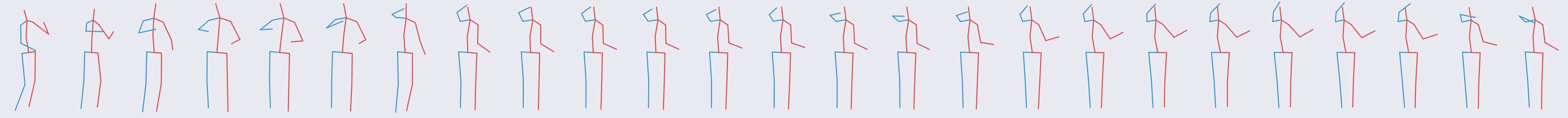}
  \end{subfigure}
  \caption{\textit{direction}: this character is directing traffic.}
  \label{fig:human_0}
\end{figure}

\begin{figure}[h!]
  \centering
  \begin{subfigure}{0.99\textwidth}
    \includegraphics[width=\textwidth]{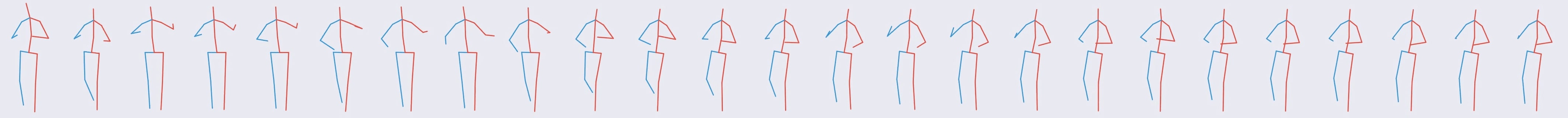}
  \end{subfigure}
  \begin{subfigure}{0.99\textwidth}
    \includegraphics[width=\textwidth]{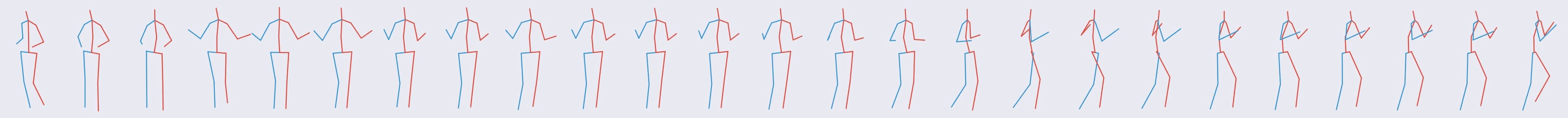}
  \end{subfigure}
  \begin{subfigure}{0.99\textwidth}
    \includegraphics[width=\textwidth]{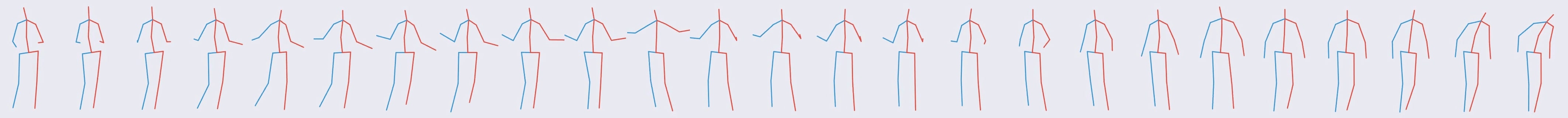}
  \end{subfigure}
  \caption{\textit{discussion}: this character is discussing issues with others.}
  \label{fig:human_1}
\end{figure}

\begin{figure}[h!]
  \centering
  \begin{subfigure}{0.99\textwidth}
    \includegraphics[width=\textwidth]{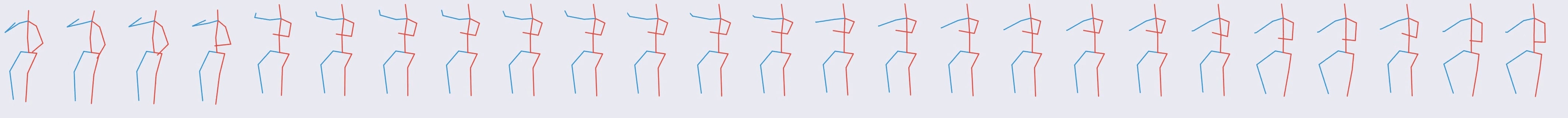}
  \end{subfigure}
  \begin{subfigure}{0.99\textwidth}
    \includegraphics[width=\textwidth]{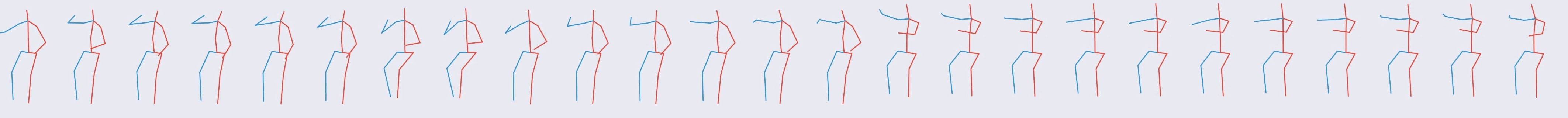}
  \end{subfigure}
  \begin{subfigure}{0.99\textwidth}
    \includegraphics[width=\textwidth]{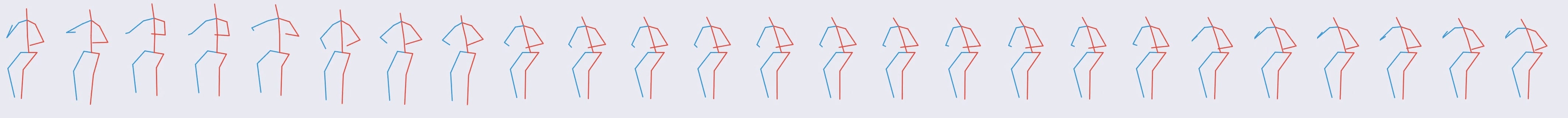}
  \end{subfigure}
  \caption{\textit{eating}: this character is sitting on the chair and having its lunch.}
  \label{fig:human_2}
\end{figure}

\begin{figure}[h!]
  \centering
  \begin{subfigure}{0.99\textwidth}
    \includegraphics[width=\textwidth]{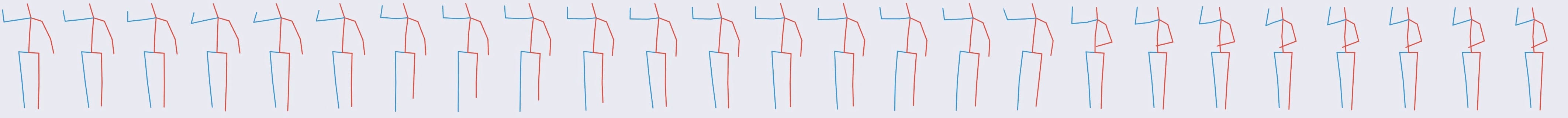}
  \end{subfigure}
  \begin{subfigure}{0.99\textwidth}
    \includegraphics[width=\textwidth]{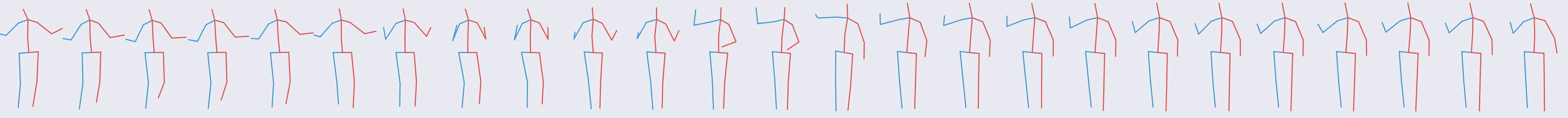}
  \end{subfigure}
  \begin{subfigure}{0.99\textwidth}
    \includegraphics[width=\textwidth]{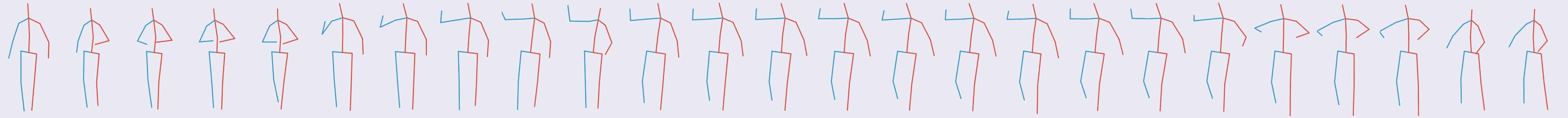}
  \end{subfigure}
  \caption{\textit{greeting}: this character is waving hands and greeting with other people.}
  \label{fig:human_3}
\end{figure}

\begin{figure}[h!]
  \centering
  \begin{subfigure}{0.99\textwidth}
    \includegraphics[width=\textwidth]{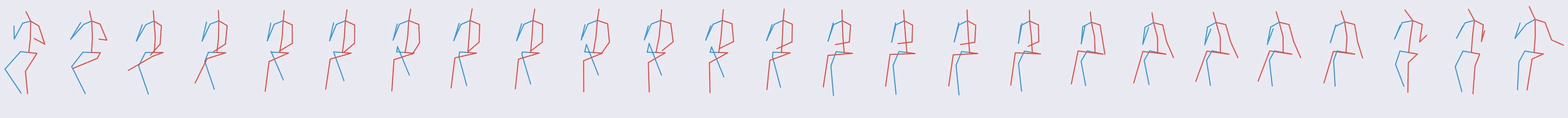}
  \end{subfigure}
  \begin{subfigure}{0.99\textwidth}
    \includegraphics[width=\textwidth]{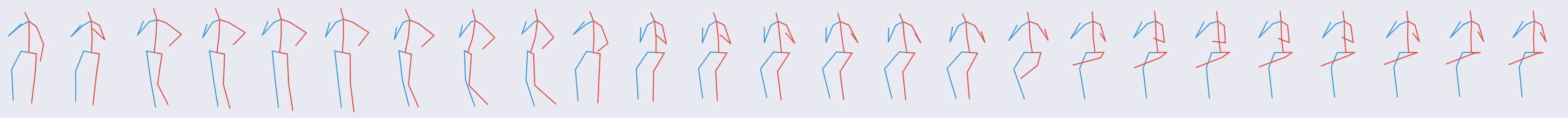}
  \end{subfigure}
  \begin{subfigure}{0.99\textwidth}
    \includegraphics[width=\textwidth]{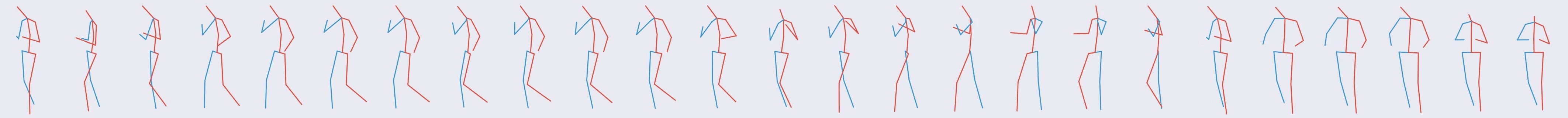}
  \end{subfigure}
  \caption{\textit{phoning}: this character is making a phone call with other people.}
  \label{fig:human_4}
\end{figure}

\begin{figure}[h!]
  \centering
  \begin{subfigure}{0.99\textwidth}
    \includegraphics[width=\textwidth]{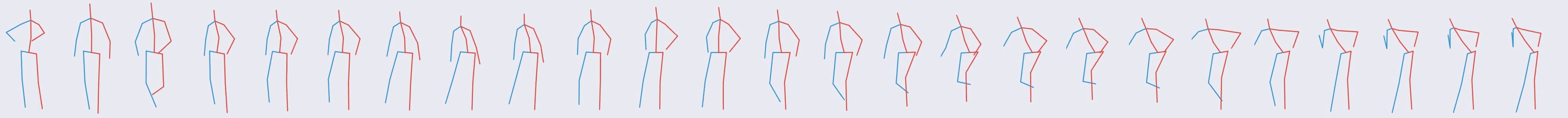}
  \end{subfigure}
  \begin{subfigure}{0.99\textwidth}
    \includegraphics[width=\textwidth]{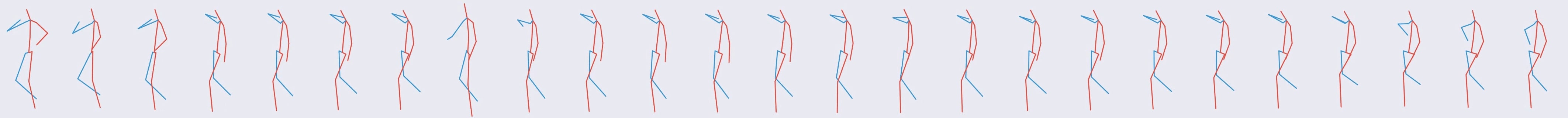}
  \end{subfigure}
  \begin{subfigure}{0.99\textwidth}
    \includegraphics[width=\textwidth]{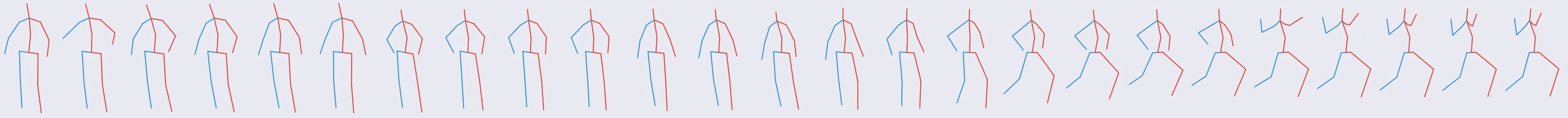}
  \end{subfigure}
  \caption{\textit{posing}: this character is making some exaggerated poses to take photos.}
\label{fig:human_5}
\end{figure}

\begin{figure}[h!]
  \centering
  \begin{subfigure}{0.99\textwidth}
    \includegraphics[width=\textwidth]{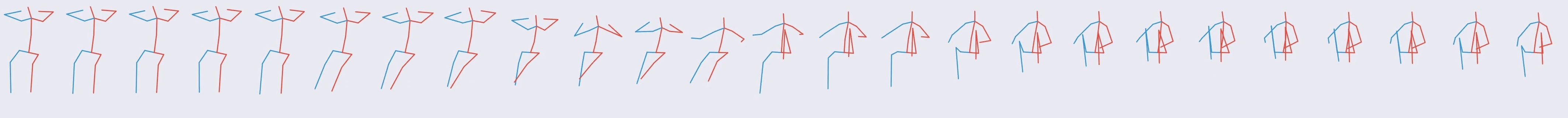}
  \end{subfigure}
  \begin{subfigure}{0.99\textwidth}
    \includegraphics[width=\textwidth]{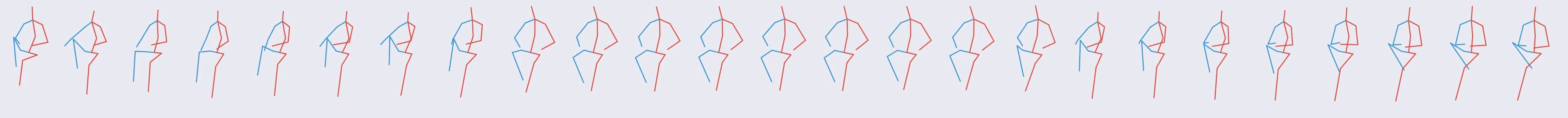}
  \end{subfigure}
  \begin{subfigure}{0.99\textwidth}
    \includegraphics[width=\textwidth]{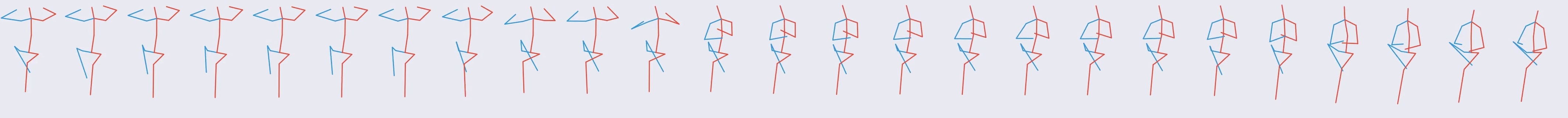}
  \end{subfigure}
  \caption{\textit{sitting}: this character is sitting down on a chair.}
  \label{fig:human_6}
\end{figure}

\begin{figure}[h!]
  \centering
  \begin{subfigure}{0.99\textwidth}
    \includegraphics[width=\textwidth]{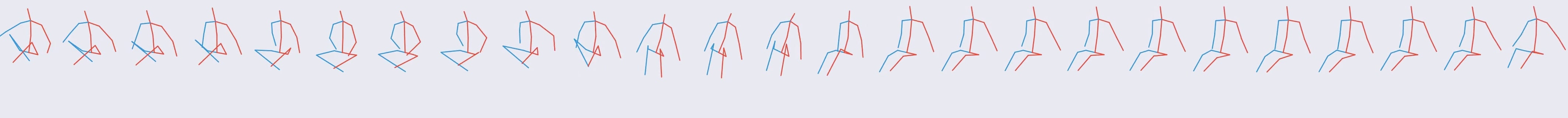}
  \end{subfigure}
  \begin{subfigure}{0.99\textwidth}
    \includegraphics[width=\textwidth]{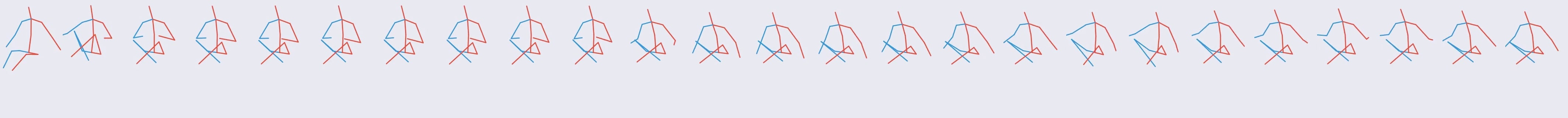}
  \end{subfigure}
  \begin{subfigure}{0.99\textwidth}
    \includegraphics[width=\textwidth]{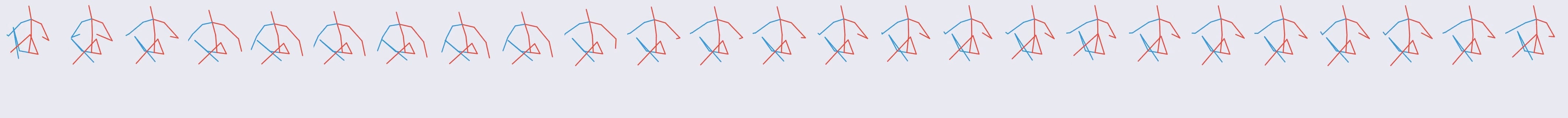}
  \end{subfigure}
  \caption{\textit{sitting down}: this character is sitting down on the ground.}
  \label{fig:human_7}
\end{figure}

\begin{figure}[h!]
  \centering
  \begin{subfigure}{0.99\textwidth}
    \includegraphics[width=\textwidth]{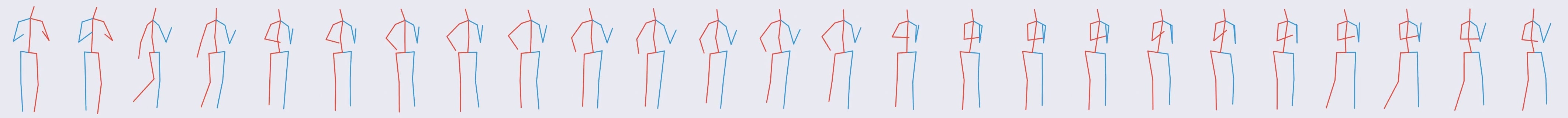}
  \end{subfigure}
  \begin{subfigure}{0.99\textwidth}
    \includegraphics[width=\textwidth]{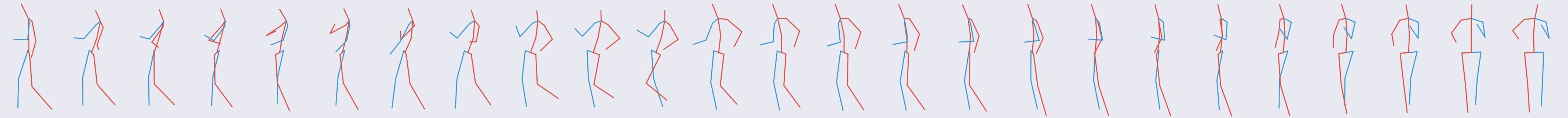}
  \end{subfigure}
  \begin{subfigure}{0.99\textwidth}
    \includegraphics[width=\textwidth]{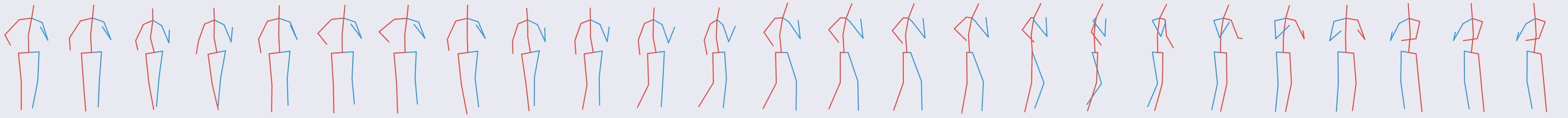}
  \end{subfigure}
  \caption{\textit{smoking}: this character is holding a cigarette in one hand and occasionally smokes.}
  \label{fig:human_8}
\end{figure}

\begin{figure}[h!]
  \centering
  \begin{subfigure}{0.99\textwidth}
    \includegraphics[width=\textwidth]{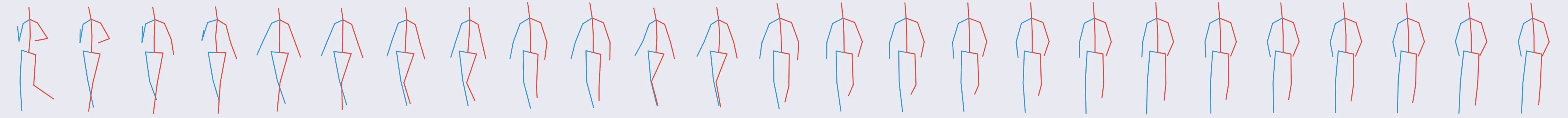}
  \end{subfigure}
  \begin{subfigure}{0.99\textwidth}
    \includegraphics[width=\textwidth]{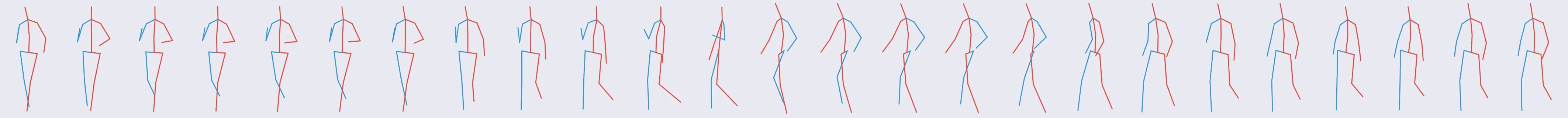}
  \end{subfigure}
  \begin{subfigure}{0.99\textwidth}
    \includegraphics[width=\textwidth]{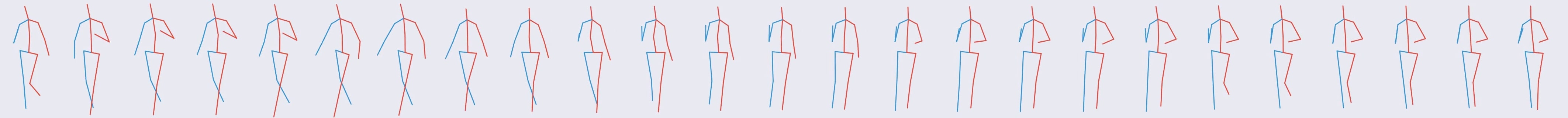}
  \end{subfigure}
  \caption{\textit{walking}: this character is walking.}
  \label{fig:human_9}
\end{figure}

\paragraph{Human-3.6m} 
We show ten classes of action sequences:
\textit{direction, discussion, eating, greeting, phoning, posing, sitting, sitting down, smoking} and \textit{walking} in Fig.~\ref{fig:human_0}, Fig.~\ref{fig:human_1}, Fig.~\ref{fig:human_2}, Fig.~\ref{fig:human_3}, Fig.~\ref{fig:human_4}, Fig.~\ref{fig:human_5}, Fig.~\ref{fig:human_6}, Fig.~\ref{fig:human_7}, Fig.~\ref{fig:human_8} and Fig.~\ref{fig:human_9} on Human-3.6m dataset. For each action class, we present three generated action sequences from random initialization.

\begin{figure}[h!]
  \centering
  \begin{subfigure}{0.99\textwidth}
    \includegraphics[width=\textwidth]{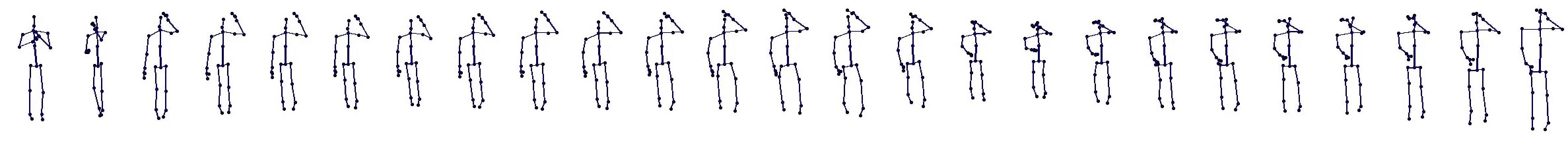}
  \end{subfigure}
  \begin{subfigure}{0.99\textwidth}
    \includegraphics[width=\textwidth]{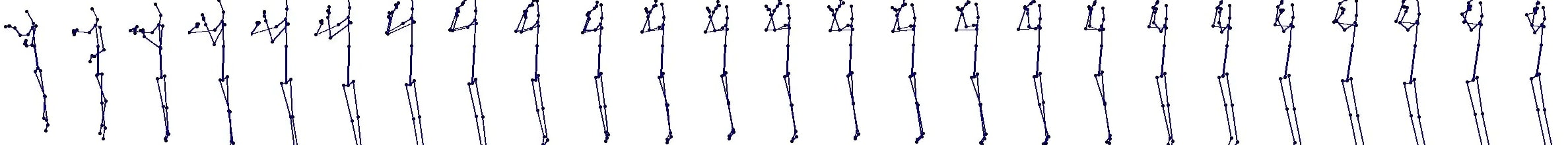}
  \end{subfigure}
  \begin{subfigure}{0.99\textwidth}
    \includegraphics[width=\textwidth]{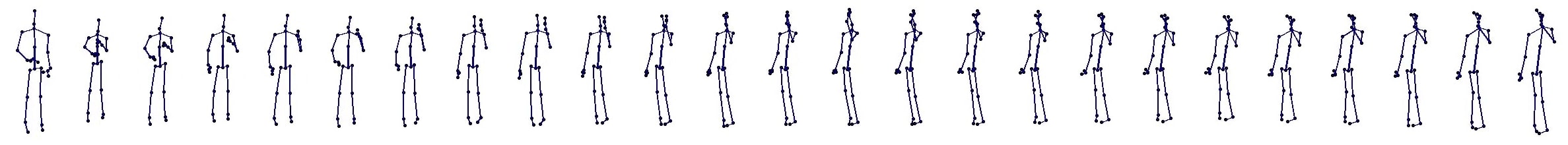}
  \end{subfigure}
  \begin{subfigure}{0.99\textwidth}
    \includegraphics[width=\textwidth]{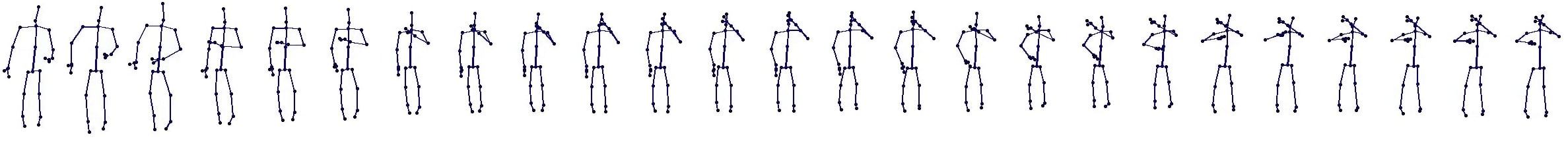}
  \end{subfigure}
  \caption{\textit{drinking water}: this character is holding a water bottle in one hand while drinking water.}
  \label{fig:ntu_0}
\end{figure}

\begin{figure}[h!]
  \centering
  \begin{subfigure}{0.99\textwidth}
    \includegraphics[width=\textwidth]{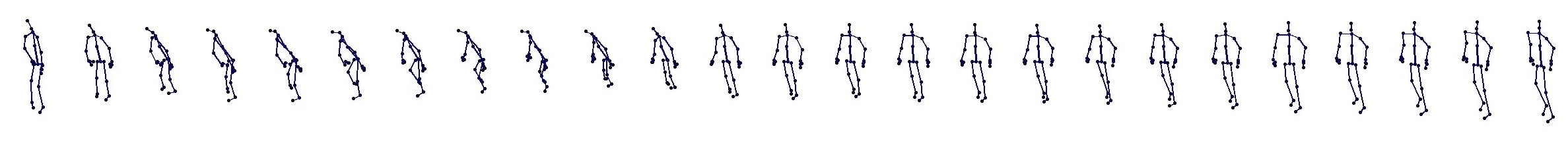}
  \end{subfigure}
  \begin{subfigure}{0.99\textwidth}
    \includegraphics[width=\textwidth]{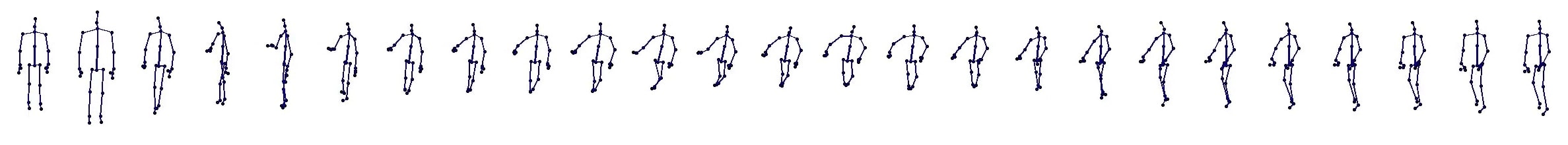}
  \end{subfigure}
  \begin{subfigure}{0.99\textwidth}
    \includegraphics[width=\textwidth]{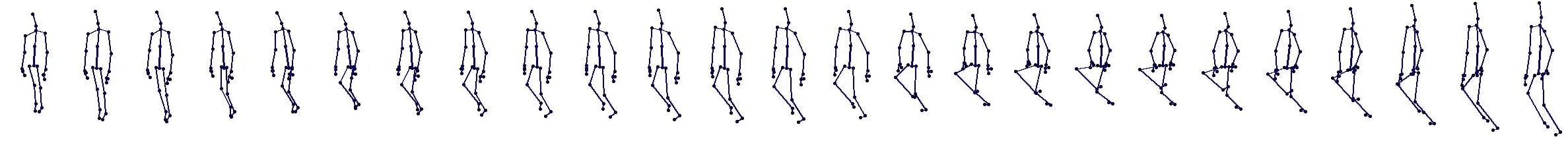}
  \end{subfigure}
  \begin{subfigure}{0.99\textwidth}
    \includegraphics[width=\textwidth]{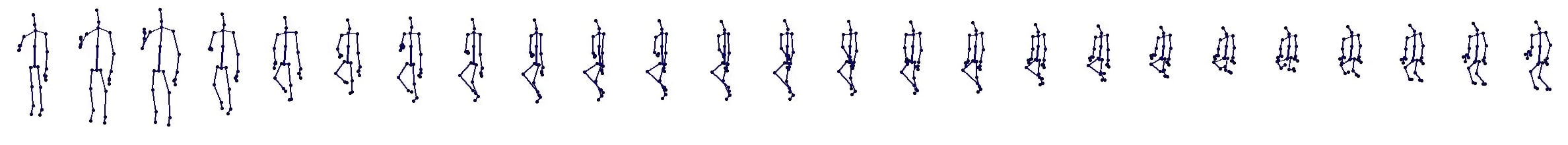}
  \end{subfigure}
  \caption{\textit{jumping up}: this character is jumping.}
  \label{fig:ntu_1}
\end{figure}

\begin{figure}[h!]
  \centering
  \begin{subfigure}{0.99\textwidth}
    \includegraphics[width=\textwidth]{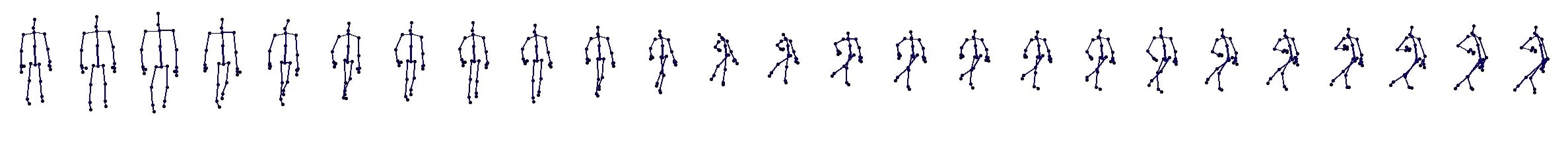}
  \end{subfigure}
  \begin{subfigure}{0.99\textwidth}
    \includegraphics[width=\textwidth]{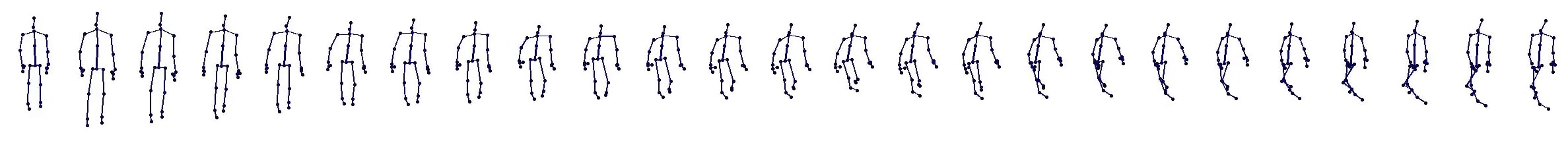}
  \end{subfigure}
  \begin{subfigure}{0.99\textwidth}
    \includegraphics[width=\textwidth]{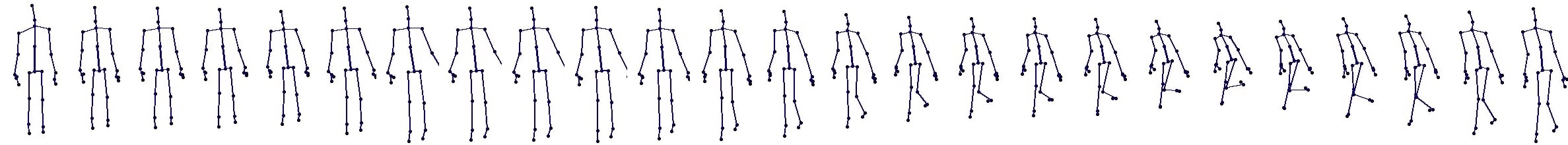}
  \end{subfigure}
  \begin{subfigure}{0.99\textwidth}
    \includegraphics[width=\textwidth]{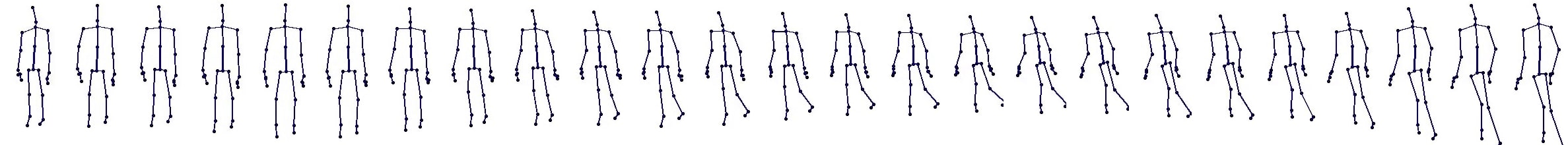}
  \end{subfigure}
  \caption{\textit{kicking something}: this character is kicking something.}
  \label{fig:ntu_2}
\end{figure}

\begin{figure}[h!]
  \centering
  \begin{subfigure}{0.99\textwidth}
    \includegraphics[width=\textwidth]{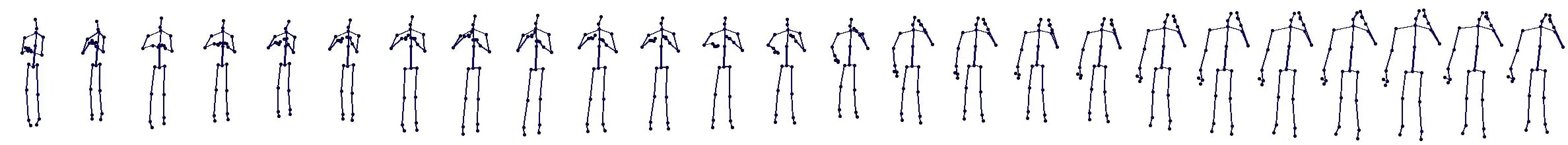}
  \end{subfigure}
  \begin{subfigure}{0.99\textwidth}
    \includegraphics[width=\textwidth]{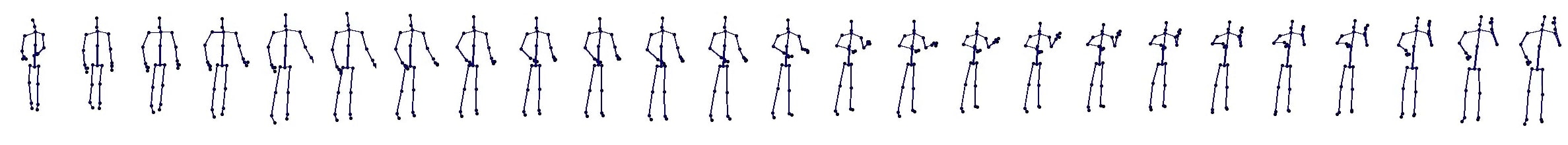}
  \end{subfigure}
  \begin{subfigure}{0.99\textwidth}
    \includegraphics[width=\textwidth]{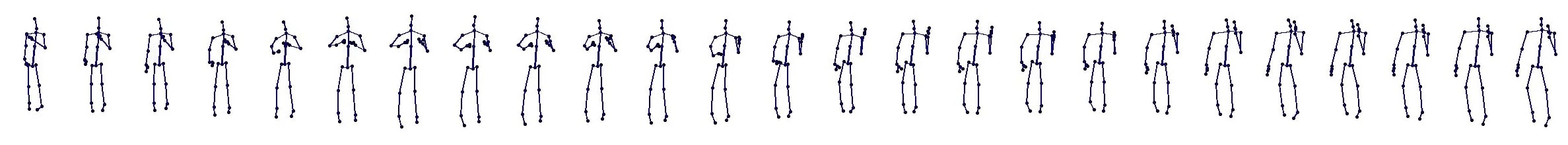}
  \end{subfigure}
  \begin{subfigure}{0.99\textwidth}
    \includegraphics[width=\textwidth]{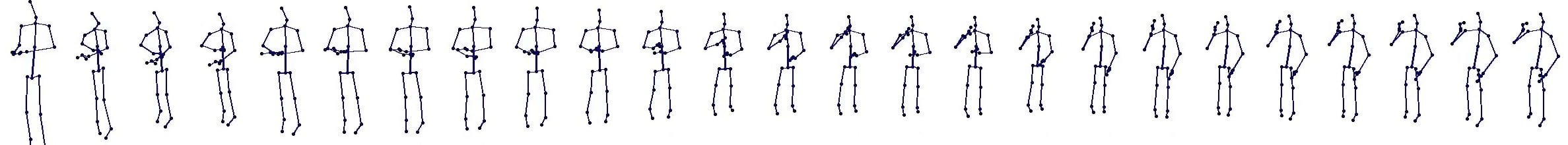}
  \end{subfigure}
  \caption{\textit{making phone call}: this character is raising his mobile phone with one hand and is making a phone call.}
  \label{fig:ntu_3}
\end{figure}

\begin{figure}[h!]
  \centering
  \begin{subfigure}{0.99\textwidth}
    \includegraphics[width=\textwidth]{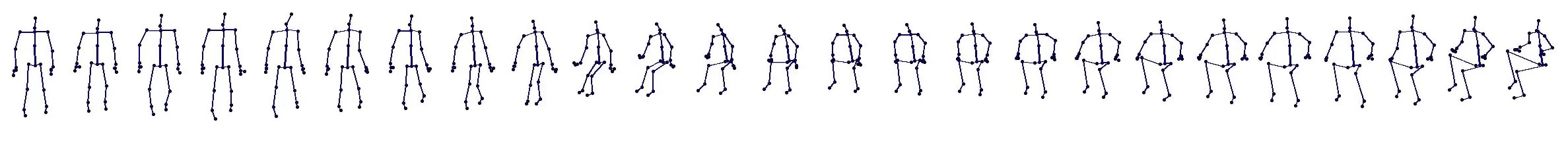}
  \end{subfigure}
  \begin{subfigure}{0.99\textwidth}
    \includegraphics[width=\textwidth]{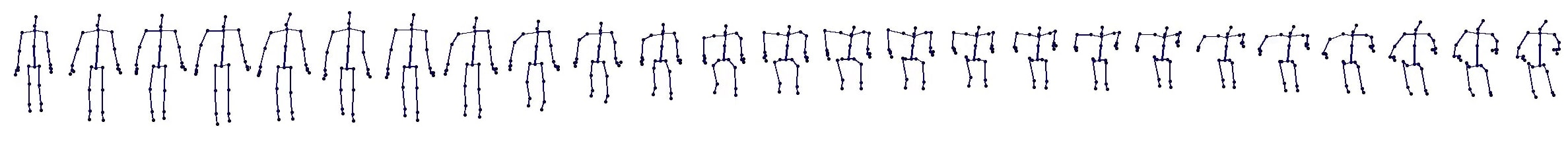}
  \end{subfigure}
  \begin{subfigure}{0.99\textwidth}
    \includegraphics[width=\textwidth]{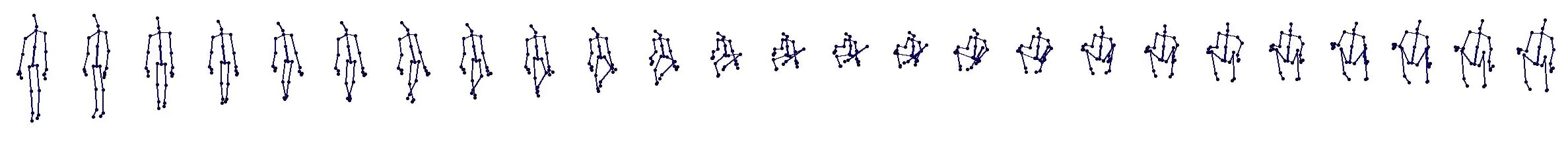}
  \end{subfigure}
  \begin{subfigure}{0.99\textwidth}
    \includegraphics[width=\textwidth]{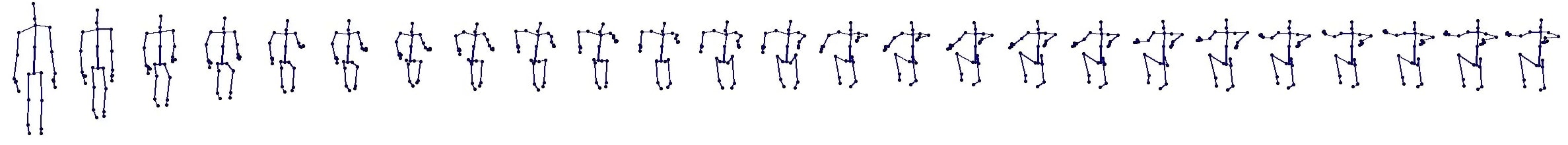}
  \end{subfigure}
  \caption{\textit{sitting down}: this character is sitting down.}
  \label{fig:ntu_4}
\end{figure}

\begin{figure}[h!]
  \centering
  \begin{subfigure}{0.99\textwidth}
    \includegraphics[width=\textwidth]{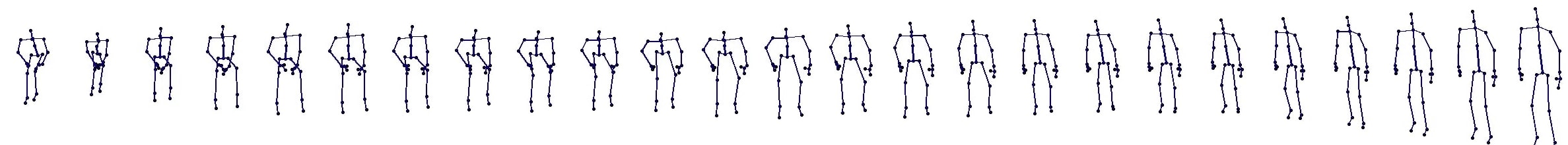}
  \end{subfigure}
  \begin{subfigure}{0.99\textwidth}
    \includegraphics[width=\textwidth]{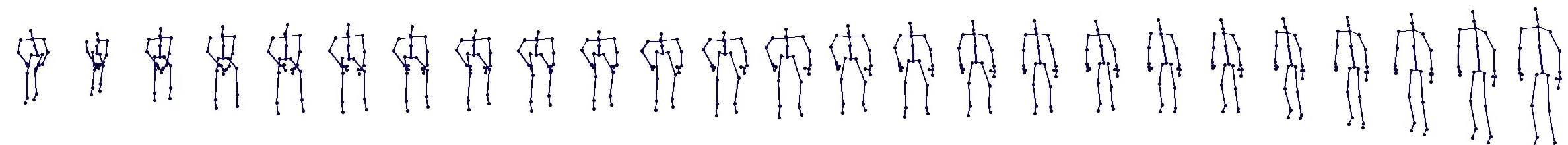}
  \end{subfigure}
  \begin{subfigure}{0.99\textwidth}
    \includegraphics[width=\textwidth]{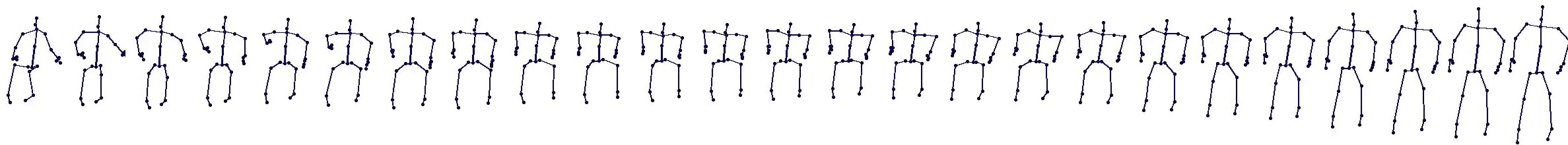}
  \end{subfigure}
  \begin{subfigure}{0.99\textwidth}
    \includegraphics[width=\textwidth]{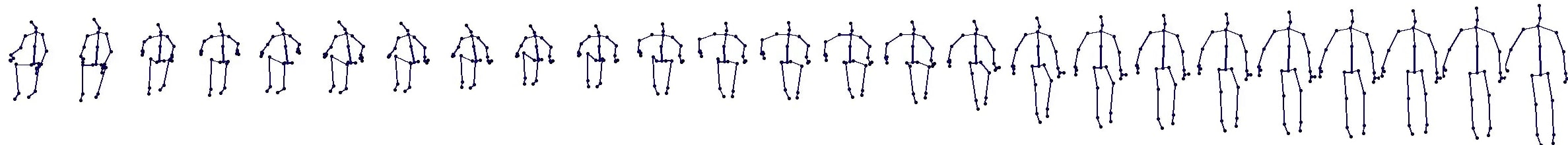}
  \end{subfigure}
  \caption{\textit{standing up}: this character is standing up.}
  \label{fig:ntu_5}
\end{figure}

\begin{figure}[h!]
  \centering
  \begin{subfigure}{0.99\textwidth}
    \includegraphics[width=\textwidth]{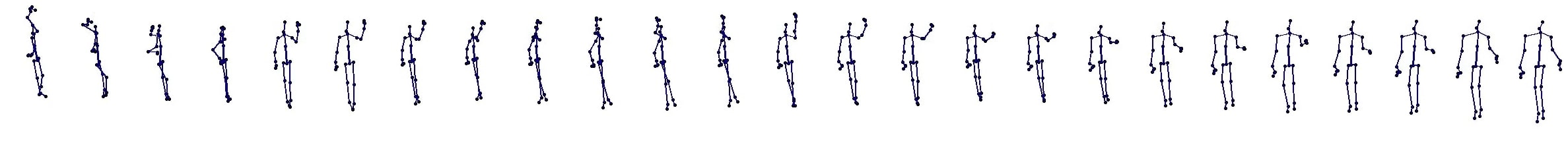}
  \end{subfigure}
  \begin{subfigure}{0.99\textwidth}
    \includegraphics[width=\textwidth]{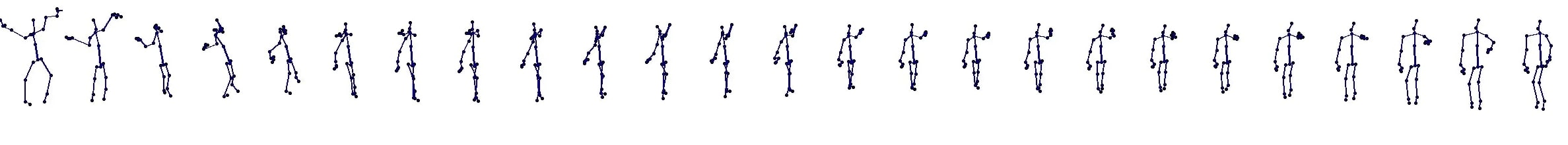}
  \end{subfigure}
  \begin{subfigure}{0.99\textwidth}
    \includegraphics[width=\textwidth]{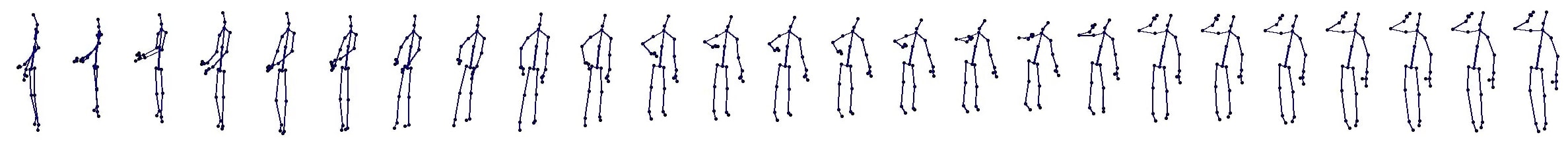}
  \end{subfigure}
  \begin{subfigure}{0.99\textwidth}
    \includegraphics[width=\textwidth]{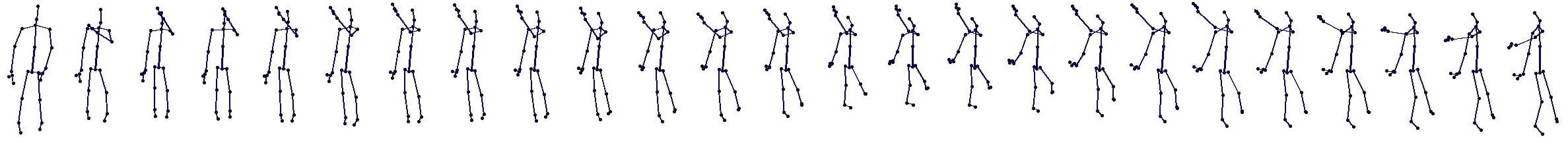}
  \end{subfigure}
  \caption{\textit{throwing}: this character is throwing a ball.}
  \label{fig:ntu_6}
\end{figure}

\begin{figure}[h!]
  \centering
  \begin{subfigure}{0.99\textwidth}
    \includegraphics[width=\textwidth]{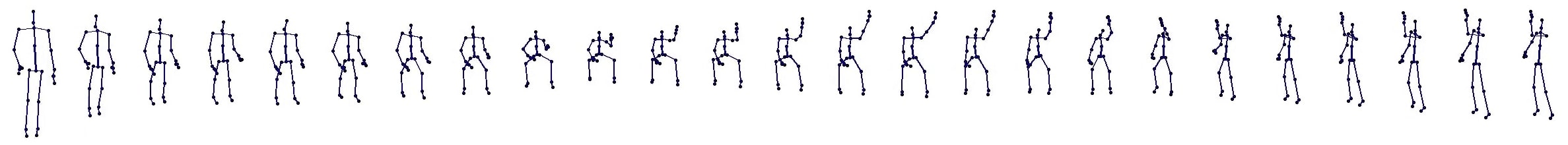}
  \end{subfigure}
  \begin{subfigure}{0.99\textwidth}
    \includegraphics[width=\textwidth]{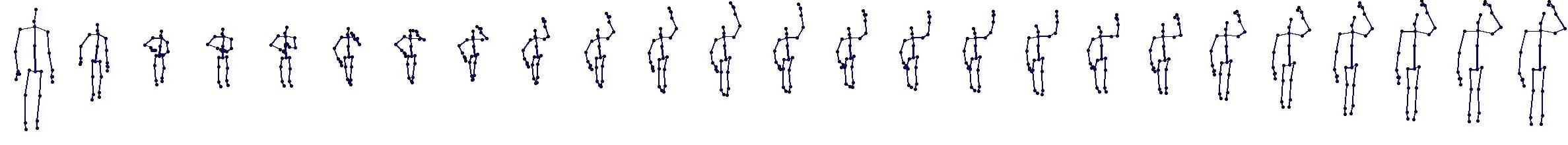}
  \end{subfigure}
  \begin{subfigure}{0.99\textwidth}
    \includegraphics[width=\textwidth]{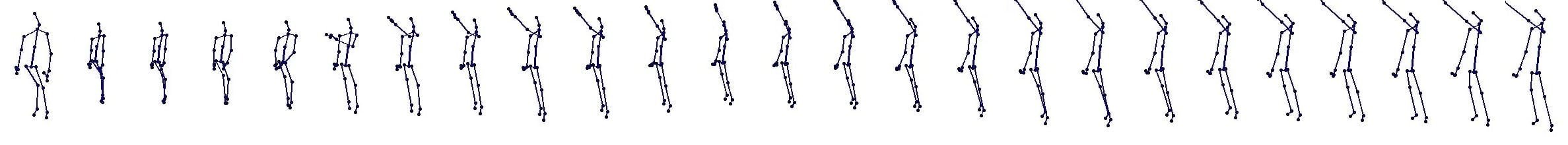}
  \end{subfigure}
  \begin{subfigure}{0.99\textwidth}
    \includegraphics[width=\textwidth]{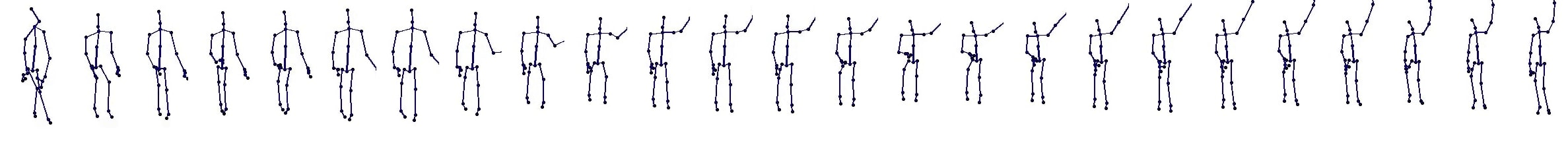}
  \end{subfigure}
  \caption{\textit{hand waving}: this character is waving hands.}
  \label{fig:ntu_7}
\end{figure}

\begin{figure}[h!]
  \centering
  \begin{subfigure}{0.99\textwidth}
    \includegraphics[width=\textwidth]{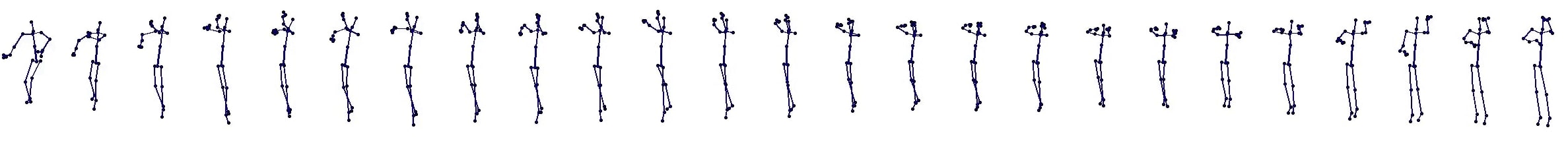}
  \end{subfigure}
  \begin{subfigure}{0.99\textwidth}
    \includegraphics[width=\textwidth]{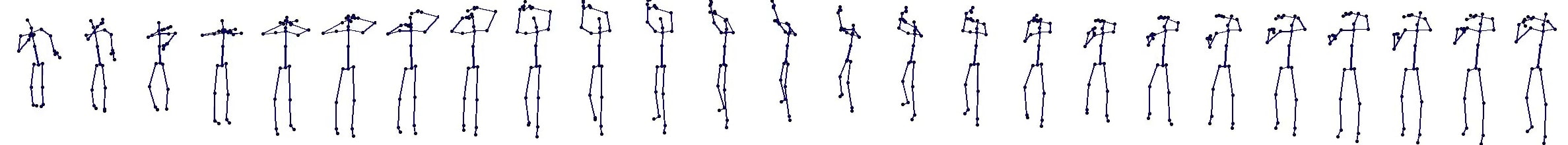}
  \end{subfigure}
  \begin{subfigure}{0.99\textwidth}
    \includegraphics[width=\textwidth]{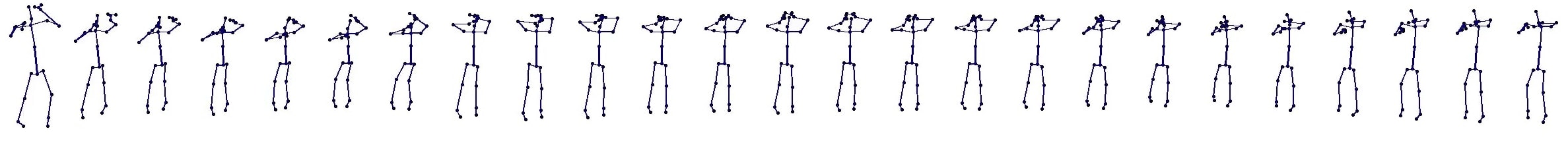}
  \end{subfigure}
  \begin{subfigure}{0.99\textwidth}
    \includegraphics[width=\textwidth]{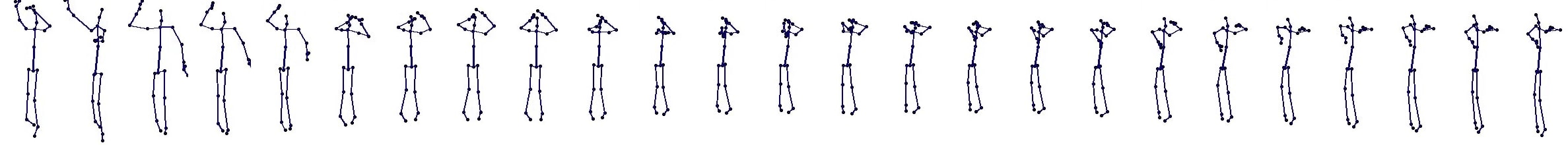}
  \end{subfigure}
  \caption{\textit{wearing jacket}: this character is wearing jacket with its two arms.}
  \label{fig:ntu_8}
\end{figure}

\begin{figure}[h!]
  \centering
  \begin{subfigure}{0.99\textwidth}
    \includegraphics[width=\textwidth]{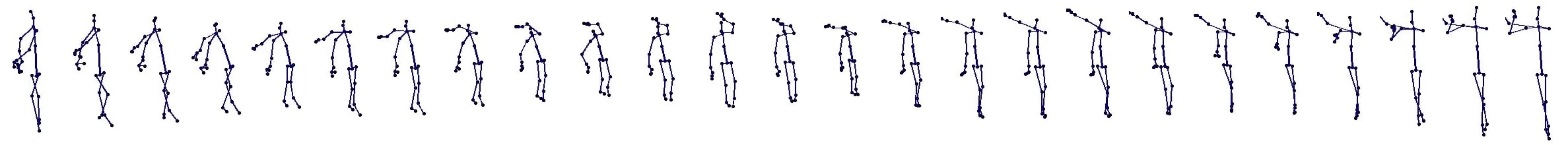}
  \end{subfigure}
  \begin{subfigure}{0.99\textwidth}
    \includegraphics[width=\textwidth]{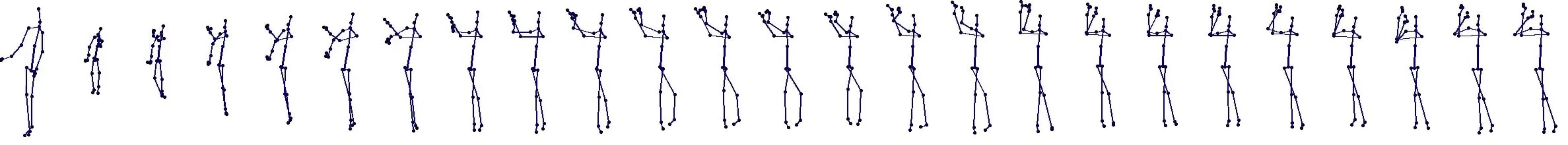}
  \end{subfigure}
  \begin{subfigure}{0.99\textwidth}
    \includegraphics[width=\textwidth]{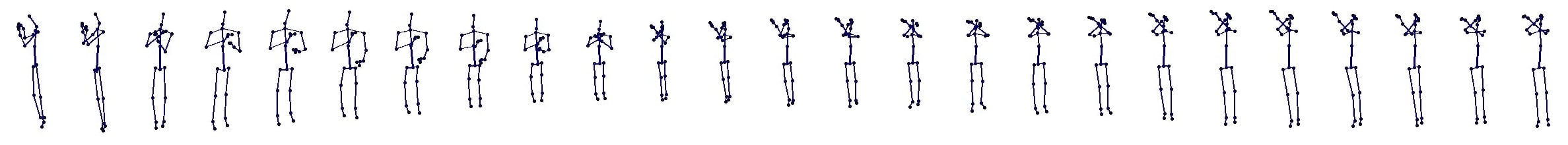}
  \end{subfigure}
  \begin{subfigure}{0.99\textwidth}
    \includegraphics[width=\textwidth]{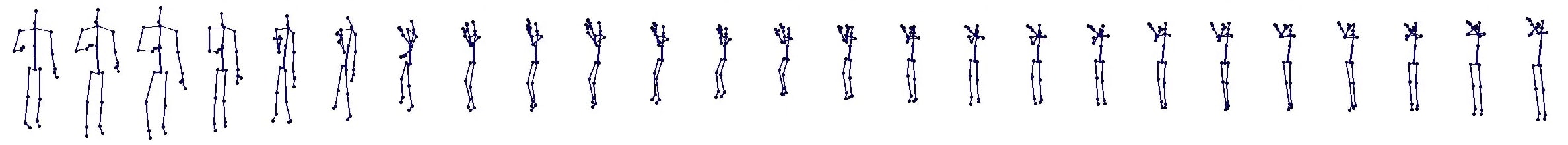}
  \end{subfigure}
  \caption{\textit{crossing hand in front}: the final position for this action is making this character's arm crossing in front.}
  \label{fig:ntu_9}
\end{figure}

\paragraph{NTU RGB+D}
We show ten classes of action sequences: \textit{drinking water, jumping up, kicking something, making phone call, sitting down, standing up, throwing, hand waving, wearing jacket} and \textit{crossing hand in front} in Fig.~\ref{fig:ntu_0}, Fig.~\ref{fig:ntu_1}, Fig.~\ref{fig:ntu_2}, Fig.~\ref{fig:ntu_3}, Fig.~\ref{fig:ntu_4}, Fig.~\ref{fig:ntu_5}, Fig.~\ref{fig:ntu_6}, Fig.~\ref{fig:ntu_7}, Fig.~\ref{fig:ntu_8} and Fig.~\ref{fig:ntu_9} on NTU RGB+D dataset. For each action class, we present two generated action sequences (the top two lines) for \textit{cross-view} and two generated action sequences (the bottom two lines) for \textit{cross-subject} from random initialization. 



\end{document}